\documentclass[prd,aps,showpacs,floats,letterpaper,floatfix,
groupedaddress,superscriptaddress,amsmath,amsfonts,
twocolumn,
,nofootinbib
,eqsecnum,
]{revtex4}
\usepackage{bm,graphics,graphicx,epsfig,color}

\usepackage{amsfonts}
\usepackage{amsopn}
\usepackage{amssymb}
\usepackage{pxfonts}
\usepackage{amssymb}
\usepackage{lscape}
\usepackage{bm}
\usepackage[varg]{txfonts}
\usepackage{color}
\usepackage{pxfonts}
\usepackage{bm,graphics,graphicx,epsfig,color}
\usepackage{ulem}
\def\be{\begin{equation}}
\def\ee{\end{equation}}

\begin{document}
\title{Parametrized tests of post-Newtonian theory using Advanced LIGO and Einstein Telescope}
\author {Chandra Kant Mishra}\email{chandra@rri.res.in}
\affiliation{Raman Research Institute, Bangalore, 560 080, India}
\affiliation{Department of Physics, Indian Institute of Science, Bangalore, 560 012, India}
\author{K. G. Arun} \email{arun@physics.wustl.edu}
\affiliation{McDonnell Center for the Space Sciences, Department of
Physics, Washington University, St.  Louis, Missouri 63130, USA}
\author {Bala R. Iyer}\email{bri@rri.res.in}
\affiliation{Raman Research Institute, Bangalore, 560 080, India}
\author{B. S. Sathyaprakash}
\email{B.Sathyaprakash@astro.cf.ac.uk}
\affiliation{School of Physics and Astronomy, Cardiff University, 
5, The Parade, Cardiff, United Kingdom, CF24 3YB}
\date{\today}
\begin{abstract}
General relativity has very specific predictions for the gravitational 
waveforms from inspiralling compact binaries obtained using the
post-Newtonian (PN) approximation. We investigate the extent to 
which the measurement of the PN coefficients, possible with the 
second generation gravitational-wave detectors such as the Advanced 
Laser Interferometer Gravitational-Wave Observatory (LIGO) and 
the third generation gravitational-wave detectors such as the Einstein 
Telescope (ET), could be used to test post-Newtonian theory and to put 
bounds on a subclass of parametrized-post-Einstein theories 
which differ from general relativity in a parametrized sense. 
We demonstrate this possibility by employing the 
best inspiralling waveform model for nonspinning compact binaries which is 3.5PN accurate in 
phase and 3PN in amplitude. Within the class of theories considered,
Advanced LIGO can test the theory at 1.5PN 
and thus the leading tail term.  Future observations of stellar mass 
black hole binaries by ET can test the consistency between the various 
PN coefficients in the gravitational-wave phasing over the mass range of 
$11$-$44 M_\odot$. The choice of the lower frequency cut off 
is  important for testing post-Newtonian theory using the ET.
The bias in the test arising from the assumption of nonspinning 
binaries is indicated.
\end{abstract}
\maketitle 
\section{Introduction}
\label{Introduction}
General relativity is tested to unprecedented accuracies in the weak-field 
and strong-field regimes (see Ref.~\cite{Will05LivRev} for a review). From 
a theoretical perspective, tests of this nature were possible due to 
physically motivated but structurally simple parametrizations of the 
observable quantities which could have  different values
 in different theories of 
gravity. In the weak-field regime, solar system bounds were largely 
assisted by the parametrized post-Newtonian (PPN) framework (see for 
example Ref.~\cite{Wthexp}). PPN formalism parametrizes the deviation of 
a general metric theory of gravity (with symmetric metric) from the 
Newtonian theory in the weak-field limit in terms of 10 free parameters 
up to order $v/c$, where $v$ is the characteristic velocity of the object. 
General relativity (GR) is a special case of this class with specific 
values of these parameters.

Binary pulsar tests, which dealt with stronger gravitational fields 
involving compact objects but typical velocities of order $v\sim10^{-3}c$, 
probed GR in the  strong-field radiative regime.  The binary pulsar tests 
were performed effectively with the use of the parametrized post-Keplerian 
(PPK)~\cite{DamourTaylor92,DD85,DD86,TW82} formulation of the pulsar 
timing formula.  The timing formula can be expressed as a function of 
Keplerian and post-Keplerian parameters, each one of which is a function 
of the component masses of the binary. A measurement of two of these 
parameters enabled the determination of the individual masses. The 
measurement of a third parameter would constitute a test of the theory, 
by requiring a consistency of the component masses in the $m_1-m_2$ plane. 
Depending on the number of these PPK parameters that can be  measured from 
the timing data of a binary pulsar, it enables many tests of GR (measuring 
$n$ parameters allow $n-2$ tests). 
Binary pulsar observations also 
confirmed the quadrupole formula for the generation of gravitational waves.

\subsection{Gravitational waves and tests of GR}

The detection of gravitational waves (GWs) would be the first direct test 
of the consistency of gravitation with the principles of special relativity 
and would probe general relativity beyond the quadrupole 
formula~\cite{SathyaSchutzLivRev09}.  A subsequent detailed study of the properties 
of GW  would next allow one to assess the validity of GR in the strong-field 
radiative regime.  A prominent class of GW sources is compact binaries: neutron 
stars (NS) and/or black holes (BH) moving in circular
orbit with velocities $v\sim0.2c$. Within GR, using different analytical and 
numerical schemes, gravitational waveforms from these systems can be computed 
with very high accuracy~\cite{Buonanno:2009zt}. Availability of such 
high-accuracy waveforms will allow the application of matched filtering 
techniques to search for these signals in the data from the GW 
interferometers such as LIGO~\cite{LIGO} and Virgo~\cite{Virgo}.  

Beyond the detection of GWs, one would like to know whether one can 
perform tests of GR with the detected signals. Despite the use of 
GR waveforms in matched filtering (which essentially assumes that 
GR is the correct theory of gravity), several authors have argued 
that GW observations can be used to test GR and put bounds on various 
parameters in alternative theories of gravity.  One of the first 
proposals towards testing nonlinear aspects of GR using GWs was due to 
Blanchet and Sathyaprakash, who discussed the possibility of measuring 
the ``tail'' effect in the GW phasing formula~\cite{BSat94,BSat95}. 
Ryan proposed a method to measure various multipole moments of a binary 
system~\cite{Ryan97} from the Laser Interferometer Space Antenna (LISA) 
observations of extreme mass ratio inspirals. Will obtained 
the additional contributions to the GW phasing formula in Brans-Dicke 
theories~\cite{Will94} and massive graviton theories ~\cite{Will98} as a 
one-parameter deviation from GR and discussed the bounds possible 
on these corresponding parameters from GW observations. These ideas
were elaborated on in greater detail in a series of papers~\cite{KKS95,WillYunes04, BBW05a, AW09,SW09,Yagi:2009zm}, studying 
how various physical effects in the binary affect the bounds. It is worth
noting that these bounds possible on massive graviton theories will be complementary to those which are obtained by binary pulsar observations (see e.g.~\cite{FinnSutton02}).

The basic idea of our proposal can be viewed as a generalization 
of some of the existing proposals to test specific theories of 
gravity like Brans-Dicke or massive graviton theories. General 
relativity and any of its parametrized variants have different 
predictions for the PN coefficients  $\psi_i$ in the phasing formula 
(for details see Sec. \ref{wformmodel}). Hence the accuracies with 
which the PN coefficients of GR can be measured translates into 
bounds on the values of these coefficients in any other theory.
This leads to the question of how well can these coefficients be 
measured?

One way to answer this question is to rephrase it as a parameter 
estimation problem and measure each of the PN coefficients, treating 
them {\it all} as independent of one another.  Recall that, for 
nonspinning binaries, each one of them is a function of only the two 
component masses and hence just two parameters are enough to describe 
the phasing formula up to any PN order. Hence, if  we want to treat 
each one of them ($8$ in all for the 
\textit{restricted waveform} (RWF) at 3.5PN order) independently, there will be
large correlations among the parameters. Our earlier work~\cite{AIQS06a}
has shown that this method works well only for a narrow range of masses
for which the signal-to-noise ratio (SNR) is high enough to discriminate these
terms. The masses are typically of a million solar mass and hence 
detectable by the LISA~\cite{AIQS06a,MSSQThesis}.

Since the high correlations among the PN parameters are responsible
for the ineffectiveness of the above test, we explored other 
possibilities which, though less generic, are viable, interesting and capture 
the essential features of the test. One possibility is to use a smaller 
set of parameters comprised of the PN coefficient to be tested 
together with any two of the remaining PN coefficients chosen as 
{\it basic variables} (to reexpress and parametrize the rest of the 
PN coefficients)~\cite{AIQS06b}.

While all these were based on inspiral waveforms, there are proposed 
tests based on merger and ringdown waveforms of the binary as well 
by measuring very accurately the
various ringdown modes of the GW spectrum~\cite{BHspect04, BCW05,HughesMenou05}. Keppel and Ajith~\cite{KeppelAjith10} revisited the bounds on massive graviton theories
including the merger and ringdown contributions. Alexander {\it et al} pointed out that
the GW observations can be used as a probe of effective quantum gravity
which predict amplitude birefringence of the spacetime for the propagation
of the GW signals~\cite{AlexanderFinnYunes08}.  Molina {\it et al} investigated the
 possible imprints of Chern-Simon theory of gravity in the
GW ringdown signals and its detectability with GW interferometers~\cite{MolinaEtAl10}.
In brief, GW measurements can
lead to interesting tests of various strong-field aspects of gravity.

Recently, Yunes and Pretorius discussed a generalized framework  called the parametrized post-Einsteinian (ppE) framework
to describe various fundamental biases in theoretical modeling of
GW and express them in a parametrized 
manner~\cite{YunesPretorius09}. They used the existing knowledge about various
alternative theories of gravity such as Brans-Dicke, massive graviton theories
and Chern-Simon theory to write down a \textit{generic} Fourier domain gravitational waveform, which is parametrized in terms of a set of amplitude and phase variables.
They also considered the contribution from  merger and ringdown phases of the binary evolution beyond the inspiral.
This parametrization in the inspiral regime can be considered  as a generalization of our earlier proposal in Ref.~\cite{AIQS06a} 
but including the possibility that amplitude of the waveform may also, 
in general, be different in an alternative theory. 

In the present work, our aim is to set up 
a general parametrization of the gravitational-wave signal in
a subclass of ppE theories that will enable 
tests of GR in the radiative regime from GW observations, similar to the 
PPN and PPK formalisms  mentioned earlier. This is an extension of 
our previous work~\cite{AIQS06b} using  more complete inspiral
waveforms, called full waveforms
(see Sec.~\ref{sec:scope} for a detailed discussion).  
From the ppE perspective, the model we have
presented in this paper would correspond to the case where there are no amplitude deviations and only one phasing coefficient 
(corresponding to the test parameter) different from GR. 
In the future the more general ppE class of meta-models must
be investigated to provide more generic results than those obtained
from the subclass we deal with in this paper.

\subsection{Choice of PN parametrization}

The  suggestion    to use a smaller 
set of parameters comprised of the PN coefficient to be tested 
together with any two of the remaining PN coefficients selected as 
{\it basic variables} (to reexpress and parametrize the rest of the 
PN coefficients)~\cite{AIQS06b} immediately raises the following question.
Which two parameters should be chosen as the basic variables? The 
most natural choice is the two lowest-order 0PN and 1PN coefficients 
since they are measured most accurately. Furthermore, within GR at 
higher PN order there are spin-orbit and spin-spin terms {\it etc}, 
so that not only the choice of higher-order coefficients as
basic parameters appears less convenient but the 
use of the lowest-order PN coefficients as basic variables may be
 expected to reduce systematic 
effects due to spins.  Once the basic variables are decided, the test 
parameter can be any of the higher PN order coefficients, chosen one at a 
time.  

The weakness of this version of the test vis-a-vis the version 
where {\it all} PN parameters are treated as independent parameters may 
be worrying at first. Unlike in the latter version where the discrepant 
PN order will be explicit in the test, in the former version the failure 
of GR at a particular PN order may not necessarily imply the corresponding 
PN term to be different from GR.  However, we believe in the robustness 
of the test itself due to the following chain of plausibility arguments: 
It is not unreasonable to assume that if an alternative theory of gravitation 
is consistent with GR at some PN order it would be normally consistent
with GR at lower PN orders but it may disagree with GR at some higher PN 
order. In view of this argument, if one is testing a particular order PN
coefficient then parametrizing the lower-order PN coefficients by the 
basic PN coefficients (at 0PN and 1PN orders) is reasonable. 

What then do we hope to achieve by our choice of expressing the 
PN coefficients of order \textit{higher} than the tested PN coefficient by the 
basic PN coefficients?
To answer this question, we make the reasonable assumption that
if a theory differs from GR at some PN order, it is likely to differ from GR 
at higher PN orders  too. 
Thus by parametrizing   
PN coefficients of orders \textit{higher} than the tested PN coefficient by the
basic PN coefficients, we naturally take into account effects in GR coming from 
higher-order PN terms and reduce the corresponding systematic errors 
arising from the higher-order PN coefficients in the 
estimation of the test parameter.  
In consequence, only departures of the \textit{correct} theory of gravity 
from GR would remain and contribute dominantly to the estimate of the PN parameter
tested in the analysis.  
To investigate this question explicitly, we computed the 
error in the estimation of a particular PN coefficient both at the
3.5PN accuracy and by truncating the expansion at the PN order of the 
test parameter. As expected,  the two choices gave 
different results, with the full phasing yielding  a more accurate 
estimation of parameters (for  more details,  see Sec.~\ref{sec:PNsys}).
Given the possibly small differences we are trying to explore any
reduction in systematic errors is indeed to be taken advantage of.

It is interesting to note that, as pointed out in Ref.~\cite{CannellaVertex09}, 
the errors in the various PN coefficients we quote here can be translated 
into measurement of three- and four-graviton vertices.  Keeping these 
caveats in view, let us consider some hypothetical theory of gravity 
which shows deviation from GR starting from  2PN order\footnote{There 
are theories which show deviation from GR starting from 2PN order. See 
for example Ref.~\cite{SopuertaYunes09} which considers one such example, 
though for spinning BHs.}.  In our proposed test, this deviation would 
not show up when $\psi_3$ is used as test parameter, as in this case the 
deviations are only from the fact that the functional dependences of
higher-order phasing coefficients on $\psi_0$ and $\psi_2$ are not the 
same as in GR. This seems less important than a lower PN order test 
parameter itself deviating from its GR value.  On the other hand, when 
$\psi_4$ or higher PN order phasing coefficients are used as test 
parameters, this deviation should be evident.  Thus, proceeding 
systematically to higher PN orders, one can ascertain the PN order 
where the new theory begins to deviate from GR.  

\subsection{Scope of the current work and a summary of results}\label{sec:scope}
In this work we revisit the problem in the context of the second 
generation ground-based GW interferometer  such as Advanced LIGO and 
a third generation ground-based GW interferometer called Einstein 
Telescope (ET) that is currently under design study in Europe.
Since ET is envisaged to have far better low-frequency sensitivity 
than Advanced LIGO (a lower frequency cutoff of about 1-5 Hz), one 
of the aims of the present investigation is to evaluate
the possible gains in going from a lower cutoff of 10 to 1 Hz. 
A further new ingredient in the present version of the above test 
is that we use not just the 3.5PN RWF but also the
amplitude-corrected full waveforms (FWF) which are 3PN accurate
in amplitude (thus having seven harmonics other than just the leading 
quadrupolar one) and 3.5PN accurate in phase.  
For nonspinning binaries, the amplitude corrections are functions of the 
two masses and the inclination angle of the binary.  
The amplitude corrections at every PN order bring new dependences on 
the binary masses and hence could improve the estimation of the phasing 
coefficients.  
With our previous insight in the weakening of the test due
to the use of more  parameters~\cite{AIQS06a}, in the first instance,
we make  the reasonable  assumption that
since  matched filtering procedure is more
sensitive to the phase rather than amplitude, one
can skip parametrizing the amplitude \textit{independently}
in the present work.
Thus the deviations in the amplitude corrections of the waveform 
are not independently parametrized  
even though they have been taken account of in the present work to
reduce systematic effects.
We rewrite the mass 
dependences in the PN amplitude terms in terms of $\psi_0$ and $\psi_2$, 
just as in the case of the phase terms. 

An obvious limitation
in regard to the present analysis concerns the 
generality of the  parametrization that we employ. Indeed, in the strong-field regime, alternative theories may so qualitatively deviate from GR
that the structure of the waveform used here for the parametrization may 
not be generic enough to capture those features 
and one may need to redo the present analysis within a more
general class of models like the ppE framework. 
It is worth mentioning that at present very accurate high PN order 
GW phasing results are available only for GR. Most results for
alternative theories of gravitation are available only for 
leading or next to leading PN orders and hence more work is required
before they are comparable in performance to the GR waveforms used
presently for GW data analysis. 
Given these circumstances, our purpose is to explore what best we can
extract from a subclass of PN models close to GR and leave to future
works a more complete investigation using e.g. ppE models.

Based on the above analysis, within the subclass of ppE theories
that we consider, we find that GW observations by advanced LIGO
of binary black holes (BBHs) in the range $11$-$110 M_{\odot}$ and 
at a luminosity distance of 300 Mpc would allow the measurement of the PN 
coefficient $\psi_3$ with fractional accuracies better that 6\% when the FWF 
is used.  On the other hand using the FWF as a waveform model and a low-frequency 
cutoff of 1 Hz, observations of stellar mass BBHs in ET would allow the 
measurement of {\it all} PN parameters, except $\psi_4$, with
accuracies better than 2\% when the total mass of the binary is in the range 
$11$-$44 M_{\odot}$.  ET observations of intermediate mass BBHs would
allow only two of the seven PN coefficients ($\psi_3$ and $\psi_{5l}$) to be
measured with fractional accuracies better than 10\% in the mass range
$55$-$400 M_{\odot}$.

The choice of a low-frequency cutoff of 1 Hz, as compared to 10 Hz, 
reduces  the relative errors in various parameters roughly by factors of 
order 2 to 10 for  stellar mass black hole binaries.  For intermediate mass 
binaries, which coalesce at smaller frequencies, though a  lower cutoff 
helps  improve the parameter estimation, the  errors associated with the 
measurement of various parameters is so large that the test is not very 
interesting.  Although the use of FWF has no particular advantage in the
case of stellar mass black hole binaries, their use in the case of 
intermediate mass black hole binaries improves parameter estimation by 
a factor of a few to almost 80.  Large improvements 
are obtained for binaries that are more massive than $\sim 100 M_{\odot}.$ 
Error in the estimation of the various PN parameters for such systems is already so 
great that the improvement brought about by the use of FWF is not 
useful for the type of tests discussed in this study. 
For  reasonable detection rates, intermediate mass BH binaries  are at
distances greater than 3 Gpc as opposed to 300 Mpc in the stellar mass case (see,
e.g., Ref.~\cite{MandelIMBHET09}): 
The gain due to the use of FWF is offset by an order-of-magnitude loss
since sources are farther away.

The rest of this paper is organized in the following way: In Sec. 
\ref{TOG2-ET} we have introduced the noise curves that are employed
for Advanced LIGO and ET and the waveform model used in the present work. 
This is followed by a brief description of the Fisher matrix formalism 
that will be used to perform parameter estimation, our proposal for 
testing GR, the physical systems investigated and the implementation of 
the test.  In Sec.~\ref{Results}, we discuss our results and various  
issues related to the present work like the systematics due to higher-order 
PN terms, choice of parametrization used in the test and the effect of
choice of source location on the results.  Finally, in Sec.~\ref{Summary} 
we give a summary of our findings and future directions.     

\section{Test of GR with Advanced LIGO and the Einstein Telescope}
\label{TOG2-ET}
In this work we shall focus on the measurement of various PN coefficients in
the context of Advanced LIGO and the Einstein Telescope. One would like to 
investigate whether the observations of stellar mass black holes binaries 
in Advanced LIGO (with SNR $\sim$30) and observations of stellar mass (with SNR of few hundreds) 
as well as intermediate mass (with SNR $\sim$ 40) BBHs in ET, will 
allow us to measure some of the PN coefficients (if not all) with good 
accuracies.  

The first generation of long baseline interferometric gravitational-wave 
detectors (GEO600, LIGO and Virgo) have more or less reached their design
sensitivity and have operated for a number of years taking good science
quality data. They have shown that it is possible to build, control, and
operate highly sensitive instruments. All of these projects are now on 
the path toward building advanced versions with strain sensitivities a
factor of 10 better than their current versions. This is made possible with
research and technology in high power lasers, ultrahigh seismic isolation
systems, improved control systems, etc., that has been developed over the  
past decade. When completed around 2015-2017, advanced detectors are expected
to make routine observation of gravitational waves --- the most promising 
of all sources being the coalescence of binaries consisting of compact 
objects (see Sec. \ref{sec:systems} for expected binary 
coalescence rates).

While advanced detectors will open the gravitational window for astronomical
observations, the expected signal-to-noise ratios will not be routinely
large enough to carry out strong-field tests of GR or high precision
measurements of cosmological measurements. The worldwide gravitational-wave 
community has already begun to explore the technological development that is 
necessary to build detectors that are an order-of-magnitude better than the 
advanced instruments. The Einstein Telescope is a three-year conceptual design 
study funded by the European Commission with the goal to identify the 
challenges to mitigate gravity gradient and seismic noise in the low-frequency
region to make it possible to observe in the 1-10 Hz band. ET will be 
designed to also make an order-of-magnitude improvement in strain sensitivity
in the 10-1000 Hz band. Such a detector will be capable of making routine
observation of high-SNR events that will be useful for carrying out precisions
tests of general relativity.

\subsection{Advanced LIGO}\label{Advligo}
\begin{figure}[t!]
\includegraphics[width=0.40\textwidth,angle=0]{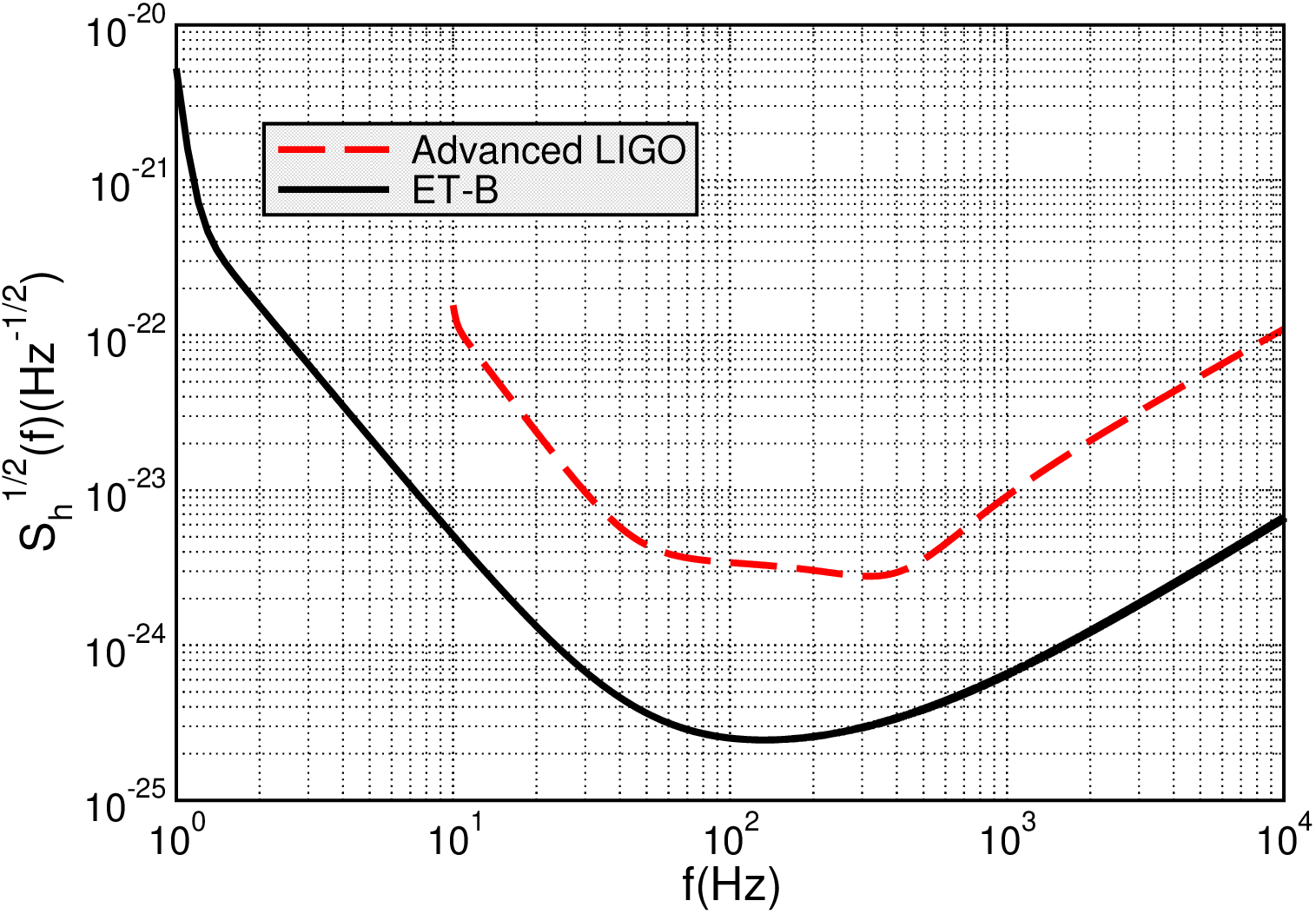}
\caption{Amplitude spectrum of Advanced LIGO and ET.}
\label{fig:noisecurves}
\end{figure}

For the studies related to Advanced LIGO we use the Advanced LIGO sensitivity 
curve~\cite{LIGO-T0900288-v3}. The analytical fit of the noise curve for Advanced 
LIGO is given by the expression,
\begin{widetext}
\begin{eqnarray}
S_h(f)&=&S_0\left[10^{16-4 (f-7.9)^2}+2.4\times10^{-62}\,x^{-50}+0.08\,x^{-4.69}
 +  123.35\,\left(\frac{1-0.23\,x^2+0.0764\,x^4}{1+0.17\,x^2}\right)\right],\; f \geq f_s\,,\nonumber\\
&=&\infty,\; f< f_s.
\label{advligo-psd}
\end{eqnarray}
\end{widetext}
where $x=f/f_0$, $f_0=215\,\mbox{Hz}$, $S_0 = 10^{-49}\,\mbox{Hz}^{-1}$ and 
$f_s$ is low-frequency cutoff below which $S_h(f)$ can be considered infinite 
for all practical purposes. We have chosen it to be 20 Hz.
The amplitude spectrum of Advanced LIGO is plotted in 
Fig.~\ref{fig:noisecurves}.

\subsection{Einstein Telescope}\label{ET}
The ET design study has come up with a number of possible sensitivity curves~\cite{ET-B}.
Examples include a single detector that operates with an improved sensitivity
over the whole band of 1 Hz to 1 kHz and a {\it xylophone} configuration consisting
of a pair of detectors, one tuned for best low-frequency (i.e., 1-100 Hz) sensitivity 
and a second detector tuned for optimal performance at higher frequencies of 100 Hz 
to a few kHz.  In our studies in this paper, we will use the {\it ET-B} sensitivity 
curve~\cite{ET-B}, which is also the official sensitivity curve for ET. 

An analytical fit to the ET sensitivity curve is given by
\label{ETB-sqrtpsd}
\begin{eqnarray}
S_h^{1/2}(f) & = & S_0^{1/2} \left[a_1 x^{b_1}+a_2 x^{b_2}
+a_3 x^{b_3}+a_4 x^{b_4}\right],\; f \geq f_s\, \nonumber\\
&=& \infty,\; f < f_s,
\end{eqnarray}
where $x=f/f_0$, $f_0=100\,\mbox{Hz}$, $S_0 = 10^{-50}\,\mbox{Hz}^{-1}$ and $f_s$ is low-frequency
cutoff below which $S_h(f)$ can be considered infinite for all practical purposes. Also one has
\begin{eqnarray}
a_1 &=&2.39\times10^{-27},~\quad
b_1=-15.64,\nonumber\\
a_2 &=&0.349, \quad\quad\quad\quad
b_2=-2.145,\nonumber\\
a_3 &=&1.76, ~~\quad\quad\quad\quad
b_3=-0.12,\nonumber\\
a_4 &=&0.409, \quad\quad\quad\quad
b_4=1.10.
\label{constants_sqrtpsd}
\end{eqnarray}
The amplitude spectrum of ET is plotted in Fig.~\ref{fig:noisecurves}.
In connection with the ET design study one of the issues to be looked into 
is the science case for going down to as low a frequency as $1$ Hz versus 
a more modest choice of $10$ Hz.

\subsection{The waveform model}\label{wformmodel}
\label{sec:waveform}
The earlier papers which discussed the tests of GR, including our own
papers~\cite{AIQS06a,AIQS06b}, assumed the so-called restricted post-Newtonian
waveform (RWF) for quasicircular, adiabatic inspiral, which contains the
dominant harmonic at twice the orbital frequency and no corrections to the
amplitude. In the present study, we include the effects of subdominant harmonics in
the waveforms. Such a wave-form, as mentioned earlier,  is called the FWF and includes
harmonics other than the dominant one, each having PN
corrections to their amplitudes. At present the most accurate waveforms
available include PN corrections in amplitude to 3PN order and in phase to
3.5PN order~\cite{BFIS08,BFIJ02,BDEI04}. To see how one might test GR or, 
more precisely, the structure of the PN theory, let us begin by considering the 
waveform from a binary in the frequency domain. The full signal in its general 
form reads as  
\begin{eqnarray}
\tilde{h}(f) &=&  \frac{2M\nu}{D_L}
\,\sum_{k=1}^{8}\,\sum_{n=0}^6\,
\frac{A_{(k,n/2)}\left (t\left (f_k \right ) \right )
\,x^{\frac{n}{2}+1}\left (t\left (f_k\right )\right )}
{2\sqrt{k\dot{F}\left (t\left (f_k\right )\right )}}
\nonumber\\ & \times &
\exp\left[{-i\phi_{(k,n/2)}\left (t\left (f_k\right )\right )} + 2 \pi i f t_c - i\pi/4 
+ i k \Psi\left (f_k \right ) \right]\,, \nonumber \\
\label {FT}
\end{eqnarray}
where $f_k=f/k,$ and the Fourier phase $\Psi(f)$ is given by
\begin{equation} 
\Psi(f) = - \phi_c + \sum_{j=0}^7\left[ \psi_j+\psi_{jl}\ln f\right] f^{(j-5)/3}\,.
\label{eq:Fourier Phase}
\end{equation}

Here $t_c$ and $\phi_c$ are the fiducial epoch of merger and the phase of the
signal at that epoch, respectively. Quantities appearing in Eq.\eqref{FT} with 
argument $t(f_k)$ denote their values at the time when the instantaneous orbital 
frequency $F(t)$ sweeps past the value $f/k$ and $x(t)$ is the PN parameter given 
by $x(t)=[2 \pi M F(t)]^{2/3}$. $A_{(k,n/2)}(t)$ and $\phi_{(k,n/2)}(t)$ are the 
polarization amplitudes and phases of the $k$th harmonic at $n/2$th PN order in 
amplitude.  The coefficients in the PN expansion of the Fourier phase are given by
\begin{eqnarray}
\psi_j &=& \frac{3}{256\,\nu}(2 \pi M)^{(j-5)/3}\alpha_j,\nonumber\\
\psi_{jl}&=&\frac{3}{256\,\nu}(2 \pi M)^{(j-5)/3}\alpha_{jl}.
\label{eq:phasingcoefficients1}
\end{eqnarray}
where,
\begin{widetext}
\allowdisplaybreaks{\begin{eqnarray}
\alpha_0 &=&1,\,\,\,  
\alpha_1 =0,\,\,\,
\alpha_2 =\frac{3715}{756}+\frac{55}{9}\nu,\,\,\, 
\alpha_3 =-16 \pi,\,\,\,
\alpha_4 =\frac{15293365}{508032}+\frac{27145}{504} \nu+\frac{3085}{72}
\nu^2; \,\,\,\alpha_{jl}=0,\, j=0,1,2,3,4,7 \nonumber\\
\alpha_5 &=&\pi \left(\frac{38645}{756}-\frac{65}{9}\nu\right)
\left[1+\ln\left(2\,6^{3/2}\pi M\right)\right], \,\,\,
\alpha_{5l} =\pi\left(\frac{38645}{756}-\frac{65}{9}\nu\right),\nonumber\\
\alpha_6 &=&\frac{11583231236531}{4694215680}-\frac{640}{3}\pi
^2-\frac{6848}{21}C
+\left(-\frac{15737765635}{3048192}+ \frac{2255}{12}\pi ^2\right)\nu
+\frac{76055}{1728}\nu^2-\frac{127825}{1296}\nu^3
-\frac{6848}{63}\ln\left(128\,\pi M\right),\nonumber\\
\alpha_{6l}&=&-\frac{6848}{63},\,\,\,
\alpha_7 =\pi\left( \frac{77096675}{254016}+\frac{378515}{1512}
\nu -\frac{74045}{756}\nu ^2\right).
\label{eq:phasingcoefficients2}
\end{eqnarray}}
\end{widetext}
The constant $C=0.577\cdots,$ appearing in the expression
for $\alpha_6,$ is Euler's constant. 

We have total nine post-Newtonian parameters: seven of these are 
the coefficients of $v^n$ terms for $n=0,2,3,4,5,6,7$ and two are 
coefficients of $v^n \ln v$ terms for $n=5,6$. These are PN coefficients 
in Einstein's theory and are functions of just two mass parameters 
chosen to be the total mass $M$ and symmetric mass ratio $\nu$.

In addition to mass dependence, the amplitude corrections also 
depend on the luminosity distance of the source to the observer and 
four additional angular parameters ($\cos\theta$, $\phi$, $\psi$, $\cos\iota$) 
related to the source location and orientation~\cite{AISSV07,ChrisAnand06b}: 
$\theta$ and $\phi$ determine the sources location, $\psi$ is the polarization 
angle and $\iota$ is the inclination angle. Once the mass dependences 
in amplitude corrections are replaced by  our fundamental pair of $\psi_0$ 
and $\psi_2$, the whole waveform can be characterized by a total ten parameters 
\begin{equation}
{\mathbf p} \equiv(\ln D_L, t_c, \phi_c, \psi_0, \psi_2, \psi_T, \cos\theta, \phi, \psi, \cos\iota)
\label{fullparameter-space}
\end{equation}

\subsection{Fisher matrix and statistical errors}
\label{sec:fisher}
We  employ the Fisher matrix approach~\cite{Finn92,FinnCh93} to see how
well we can measure these parameters. Below we briefly list  the basic 
equations of the Fisher matrix approach that we subsequently need.

Let $\tilde{\theta}^{a}$ denote the ``true values'' of the parameters 
and let $\tilde{\theta}^{a}+\Delta\theta^{a}$ be the best-fit parameters 
in the presence of some realization of the noise.  Then for large SNR, 
error in the estimation of parameters $\Delta\theta^{a}$ obeys a Gaussian 
probability distribution~\cite{Helstrom68,Wainstein,Finn92,FinnCh93} of the form
\begin{equation}
p(\Delta\theta^{a})=p^{(0)} \exp\left [-\frac{1}{2}\Gamma_{bc}\Delta\theta^{b}\Delta\theta^{c}\right ],
\label{eq:prob-dist}
\end{equation}
where $p^{(0)}$ is a normalization constant.
The quantity $\Gamma_{ab}$ appearing in Eq.\eqref{eq:prob-dist} is the 
{\it Fisher information matrix} and is given by
\begin{equation}
\Gamma_{ab}=(h_{a}\,|\,h_{b})
\label{Fishermatrix}
\end{equation}
where $h_{a}\equiv \partial h/\partial \theta^a$.
Here, $(\,|\,)$ denotes the noise weighted inner product. Given any two functions
$g$ and $h$ their inner product is defined as
\begin{equation}
(g\,|\,h)\equiv 4\,{\mbox {Re}} \int_{f_{min}}^{f_{max}}df\,
\frac{\tilde g^{*}(f)\,\tilde h(f)}{S_h(f)}.\label{eq:inner product}
\end{equation}
The integration limit $[f_{min}, f_{max}]$ is determined by both the detector
and by the nature of the signal. Each harmonic in $\tilde h(f)$ is assumed
to vanish  outside a certain frequency range. The simplest physical choice is to set the
contribution from the $k$th harmonic to the waveform zero above the frequency $k
f_{\rm lso}$, where $f_{\rm lso}$ is the orbital frequency at the last stable 
orbit~\cite{ChrisAnand06}. Since the amplitude-corrected waveform we are using 
in this work has eight harmonics, we set the upper cutoff to be 8$f_{\rm lso}$ 
when we use the FWF in the analysis. For lower cutoff, as power spectral 
densities $S_h(f)$ tend to rise very quickly below a certain frequency $f_s$ 
where they can be considered infinite for all practical purposes, we may set 
it to be $f_s$. Using the definition of the inner product one can reexpress 
$\Gamma_{ab}$ more explicitly as
\begin{equation}
\Gamma_{ab} = 4 \int_{f_s}^{k f_{\rm lso}}
\frac{\mbox {Re}(\tilde{h}_{a}^*(f) \tilde{h}_{b}(f))}{S_h(f)}\; df.\label{eq:gamma-eqn}
\end{equation}
The covariance matrix, defined as the inverse of the Fisher matrix, is given by 
\begin{equation}
\Sigma^{ab} \equiv \langle \Delta \theta^a
\Delta \theta^b \rangle = ( {\Gamma}^{-1})^{ab},
\label{sigma_a}
\end{equation}
where $\langle \cdot \rangle$ denotes an average over the 
probability distribution function in Eq.~(\ref{eq:prob-dist}). 
The root-mean-square error $\sigma_a$ in the estimation of the
parameters $\theta^{a}$ is
\begin{equation}
\sigma_a =
\bigl\langle (\Delta \theta^a)^2 \bigr\rangle^{1/2}
= \sqrt{\Sigma^{aa}}\,,
\label{eq:sigma_a}
\end{equation}
In the present work we deal with inspiralling compact binaries as seen by 
Earth bound detectors.  For such burst sources, one can  approximate the 
detector's beam pattern functions as being constant over the duration of the 
signal and thus we can assume that angular parameters ($\cos\theta$, $\phi$ 
and $\psi$) as well as the luminosity distance ($D_L$) are fixed and thus 
can be excluded from the analysis. With this restriction, the large 
10-dimensional parameter space reduces to a smaller 6-dimensional parameter 
space given by 
\begin{equation}
{\mathbf p} \equiv (t_c, \phi_c, \psi_0, \psi_2, \psi_T, \cos\iota)
\label{parameter-space}
\end{equation}    

In order to test the PN structure, one should be able to measure various PN
coefficients with good accuracy. In the present work we have assumed that the
relative error in the measurement of a parameter should be less than 10\%,
i.e. $\Delta \psi_j/\psi_j \le 0.1$, where $\Delta \psi_j$ is the error in
the estimation of the parameter $\psi_j$, in order to estimate its value in the
PN series with confidence.
\vskip 1.0cm
\begin{figure}[tbhp!]\centering
\includegraphics[width=0.48\textwidth,angle=0]{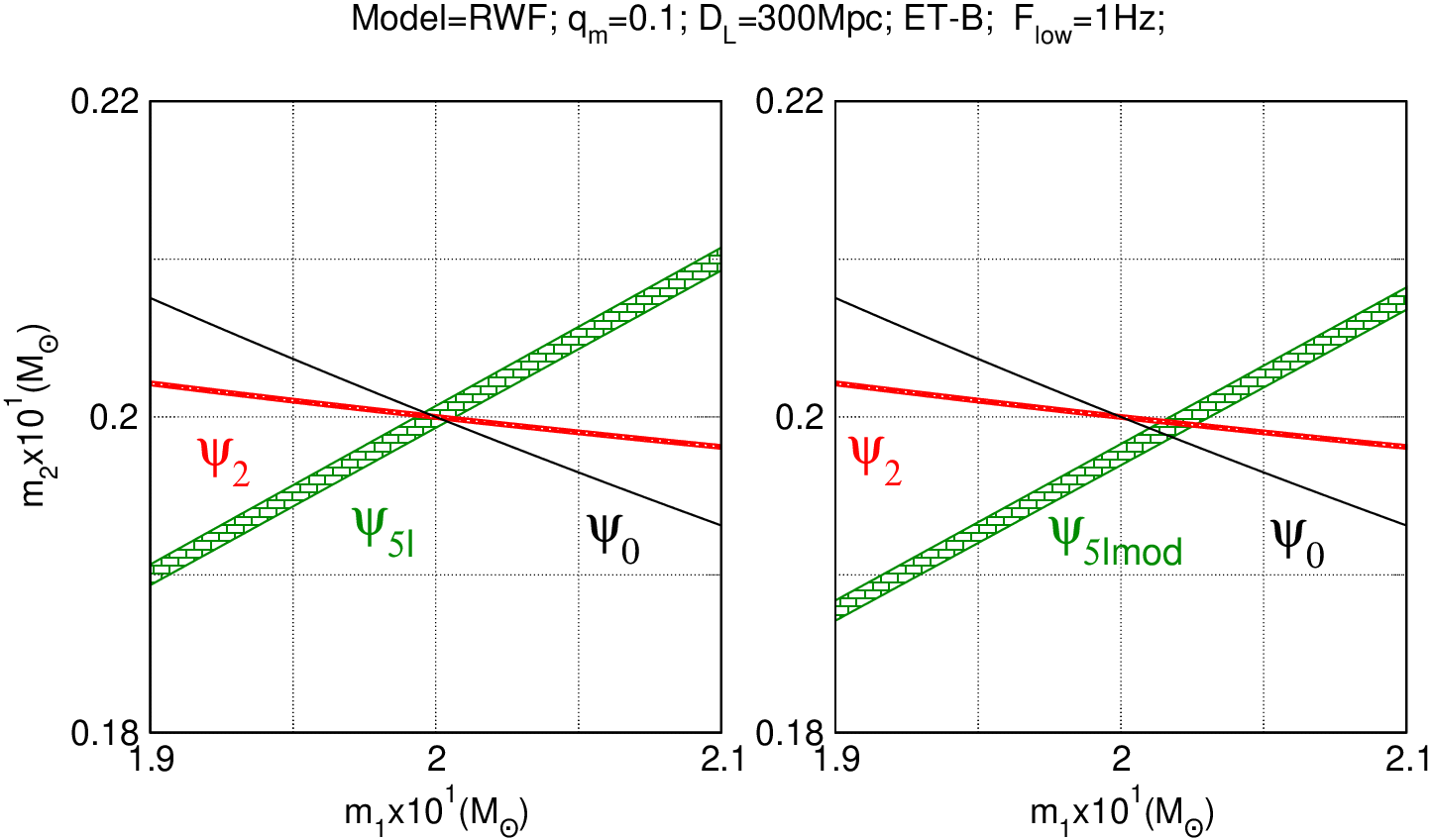}
\caption{Plots showing the regions in the $m_1$-$m_2$ plane that 
correspond to 1-$\sigma$ uncertainties in $\psi_0$, $\psi_2$ 
and $\psi_{5{\rm {l}}}$ (left panel) and those in 
$\psi_0$, $\psi_2$ and $\psi_{5{\rm {lmod}}}$ (right panel) for 
a $(2,20) M_{\odot}$ BBH at a luminosity distance of $D_{L}=300$ 
Mpc observed by ET. The low-frequency cutoff is 1Hz and RWF 
has been used. For the curves in the right panel we have assumed that the correct theory of gravity is a hypothetical non-GR theory in which  the phasing coefficient  $\psi_{5l}$ and all higher
PN coefficients differ from the GR values by 1\%.}
\label{demotog_psi5l}
\end{figure}

\subsection{Systems investigated}
\label{sec:systems}
The first detection of gravitational radiation in ground-based interferometric 
detectors is generally expected to be from the coalescence of compact binary 
systems with neutron star and black hole components~\cite{GrishRev01a}. 
Among these, binary neutron stars (BNS) are arguably the most promising ones 
with expected rates of about 40 mergers per year in Advanced LIGO and millions
of them in ET. While very interesting for other proposed tests of GR, BNS systems
are not useful for the tests proposed in this study. For our purposes a compact
binary in which one or both the components is a stellar mass ($\sim 2$-$30M_\odot$)
or intermediate mass ($\sim 50$-$1000M_\odot$) black hole (the other being a neutron
star) would be most interesting.  For our studies related to Advanced LIGO, we 
have chosen binary black holes in the mass range $11$-$110 M_\odot$ 
and their distance from the Earth to be 300 Mpc.  

For the analysis using ET we have discussed separately stellar-mass and 
intermediate-mass BBHs. For stellar mass BBHs, we have again chosen their 
luminosity distance from the Earth to be 300 Mpc and  the range of the 
total mass to be $11$-$44 M_{\odot}$. Coalescence rate of stellar 
mass BBHs is highly uncertain. The predicted rate of coalescence within a 
distance of 300 Mpc varies between one event per 10 years to several per year~ \cite{rates:2010cf}.  However, it is with such rare high-SNR events that one 
expects to perform precision tests of GR.  For intermediate mass black holes, 
we have chosen  the distance to be 3 Gpc ($z=0.55$), and their total mass to 
be in the range $55$-$1100 M_{\odot}$. The evolutionary history of 
intermediate mass BBHs and their rate of coalescence is still not well understood.  
The main motivation to study these systems comes from the models that invoke 
them as seeds of massive black holes at galactic nuclei. In a recent study, 
it has been suggested that only few coalescence events of intermediate mass 
BBHs could be expected within a redshift of $z = 2$. Also depending on what 
triggered seed galaxies there may be a few events within a redshift of $z=1$ 
~\cite{rates:2010cf,Mandel:2007hi,Sesana:2009wg,MandelIMBHET09}.

\subsection{Implementation of the test}
\label{sec:implementation}

\begin{figure*}[t]\centering
\includegraphics[width=1.0\textwidth,angle=0]{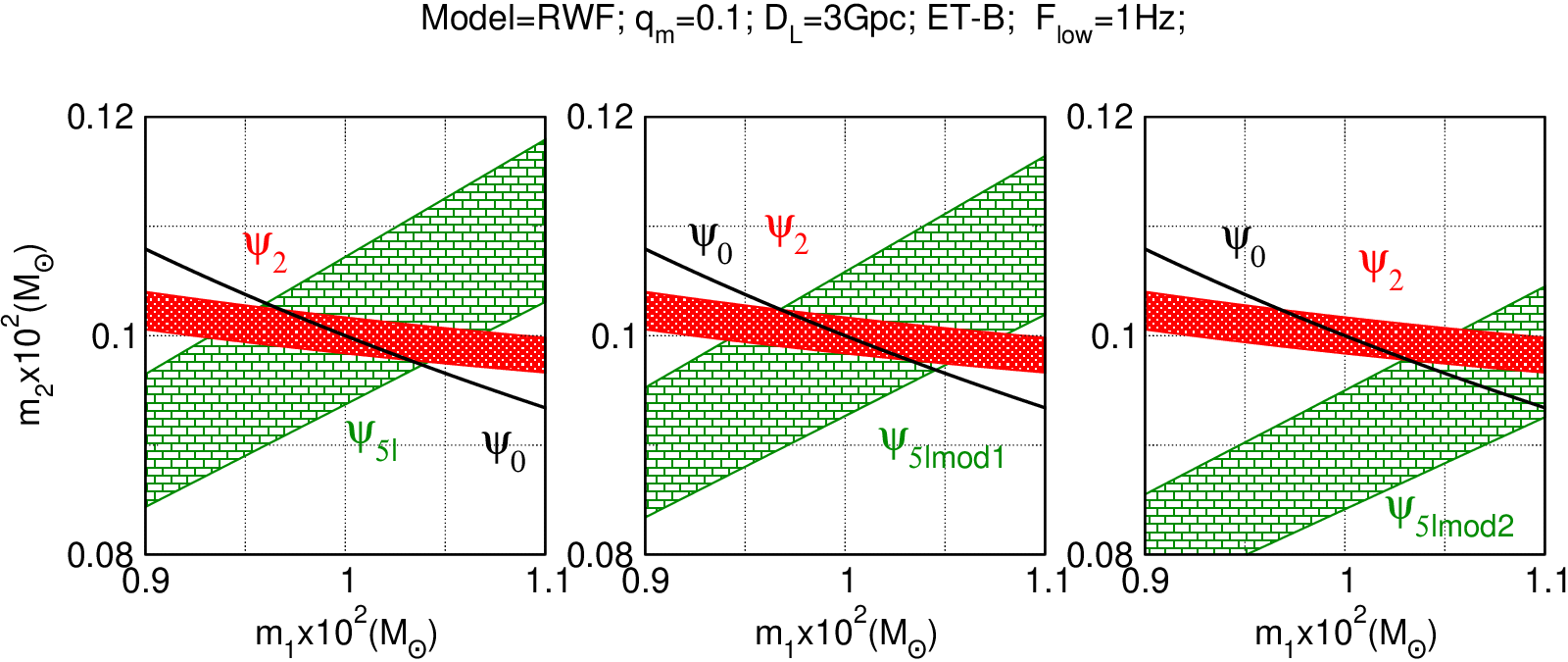}
\caption{Plots showing the regions in the $m_1$-$m_2$ plane 
that correspond to 1-$\sigma$ uncertainties in Newtonian, 1PN and 2.5PN coefficients in the PN series
for a $(10,100) M_{\odot}$ BBH at a luminosity distance of $D_{L}=3$ Gpc observed by ET. The low-frequency cutoff is 1Hz and RWF has been used.
 The left panel corresponds to GR as the correct theory of gravity while the middle and right panels correspond to hypothetical non-GR theories of gravity
 which have phasing coefficients  $\psi_{5l}$ (2.5PN) and higher  differing from the GR values by 1\% and 10\% respectively. 
}
\label{demo_tog}
\end{figure*}
\vskip 0.6cm

As mentioned earlier, in Einstein's theory (and thus in theories ``close'' 
to GR) each PN coefficient for a nonspinning compact binary is a 
function of the two mass parameters, the total mass $M$ and the symmetric 
mass ratio $\nu$. In other words, we can say that each $\psi_i$ is a 
function of the masses ($m_1$, $m_2$) of the components constituting the 
binary, i.e. $\psi_i \equiv \psi_i(m_1, m_2)$. With high-SNR 
GW observations of stellar and intermediate mass BBHs in Advanced LIGO 
and ET, it would be possible to measure the individual masses 
constituting the binary with good accuracies.  Thus, once the 
(statistical) error in the  parameter is estimated using the 
Fisher matrix, we can represent the region it spans in the space 
of masses by inverting the relation $\psi_i \equiv \psi_i(m_1, m_2)$ 
to get say $m_2 \equiv m_2(\psi_i, m_1).$ Given the measured
value $\psi_i^{\rm meas}$ and the errors $\Delta\psi_i$ in the estimation 
of $\psi_i,$ the region in the mass-plane corresponding to $m_2$ is
given by $m_2 \equiv m_2(\psi_i^{\rm meas}\pm \Delta \psi_i, m_1).$ 
For each $\psi_i$, there would be an allowed 
region in the $m_1$-$m_2$ plane and if Einstein's theory of gravity, 
or, more precisely, the PN approximation to it, is a correct theory then the 
three parameters $\psi_0$,  $\psi_2$ and $\psi_T$ (the test parameter) should have a common 
nonempty intersection in the $m_1$-$m_2$ plane. Proceeding in this way, 
for six test parameters we shall have six different tests of the theory. 
In the  present work, we shall only discuss asymmetric binaries with 
component mass ratio $q_m= 0.1$. Since the different PN coefficients 
are symmetric with respect to the exchange of $m_1$ and $m_2$, we 
expect plots in the  $m_1$-$m_2$ plane to have two symmetric branches. 
Figures~\ref{demotog_psi5l}, \ref{m1m2_22Msun} and 
~\ref{m1m2_220Msun} show one branch of the full plot.

Figure~\ref{demotog_psi5l} schematically demonstrates how the test works
by using $\psi_0$ and $\psi_2$ as basic variables and $\psi_{5l}$ as a 
test parameter.  The plot on the left uses PN coefficients predicted by 
GR, assuming GR is a correct theory of gravity.  Clearly, this shows 
that all three parameters, $\psi_0$, $\psi_2$ and $\psi_{5l},$ have 
a common nonempty intersection in the plane of masses and this 
is what we expect if  GR is the {\it correct} theory of gravity.

In contrast, consider the possibility that the correct theory of gravity is a hypothetical non-GR theory 
in which  the phasing coefficient  $\psi_{5l}$ and all higher PN coefficients,  differ from the GR values by 1\%.  
We have assumed here, in an ad-hoc manner, that the deviation of the PN terms at higher orders above the 2.5PN term (which is put to test) to be a simple scaling, i.e., $\psi_k{\rightarrow}1.01\psi_k$ for $k\geq5$. As we shall show later, there is a range of binary masses for 
which the observation of the GW signal by ET could estimate this 
coefficient with an accuracy much better than 1\%. In this 
scenario, if we interpret the $\psi_{\rm 5l}$ obtained by fitting 
to the observed GW signal, as a GR coefficient,
there will definitely be an inconsistency in the $m_1$-$m_2$ plane.  
This can clearly be  seen in the right panel of 
Fig.~\ref{demotog_psi5l} where there is no overlapping region
between the three parameters in question in the $m_1$-$m_2$ plane,
thus demonstrating the spirit of the proposed test.

Figure~\ref{demo_tog} shows a similar exercise for the 
(10, 100)$M_{\odot}$ BBH located at a luminosity distance of 
$D_L=3$Gpc observed by ET. The low-frequency cutoff is 1Hz 
and RWF has been used. The left panel in Fig.~\ref{demo_tog} 
assumes that GR is the correct theory of gravity, whereas the 
middle and right panels assume that the correct theory of 
gravity is a hypothetical non-GR theory in which the PN coefficients 
at 2.5PN (i.e., $\psi_{5l}$) and all higher orders differ from 
their GR values by 1\% and 10\% respectively.  The 2.5PN 
coefficient in GR and in the above two hypothetical theories  
can be measured with fractional accuracies of 5.6\%, 5.5\% 
and 5.1\%, respectively, for the system under consideration.

\begin{figure*}[t]
\includegraphics[width=0.42\textwidth,angle=0]{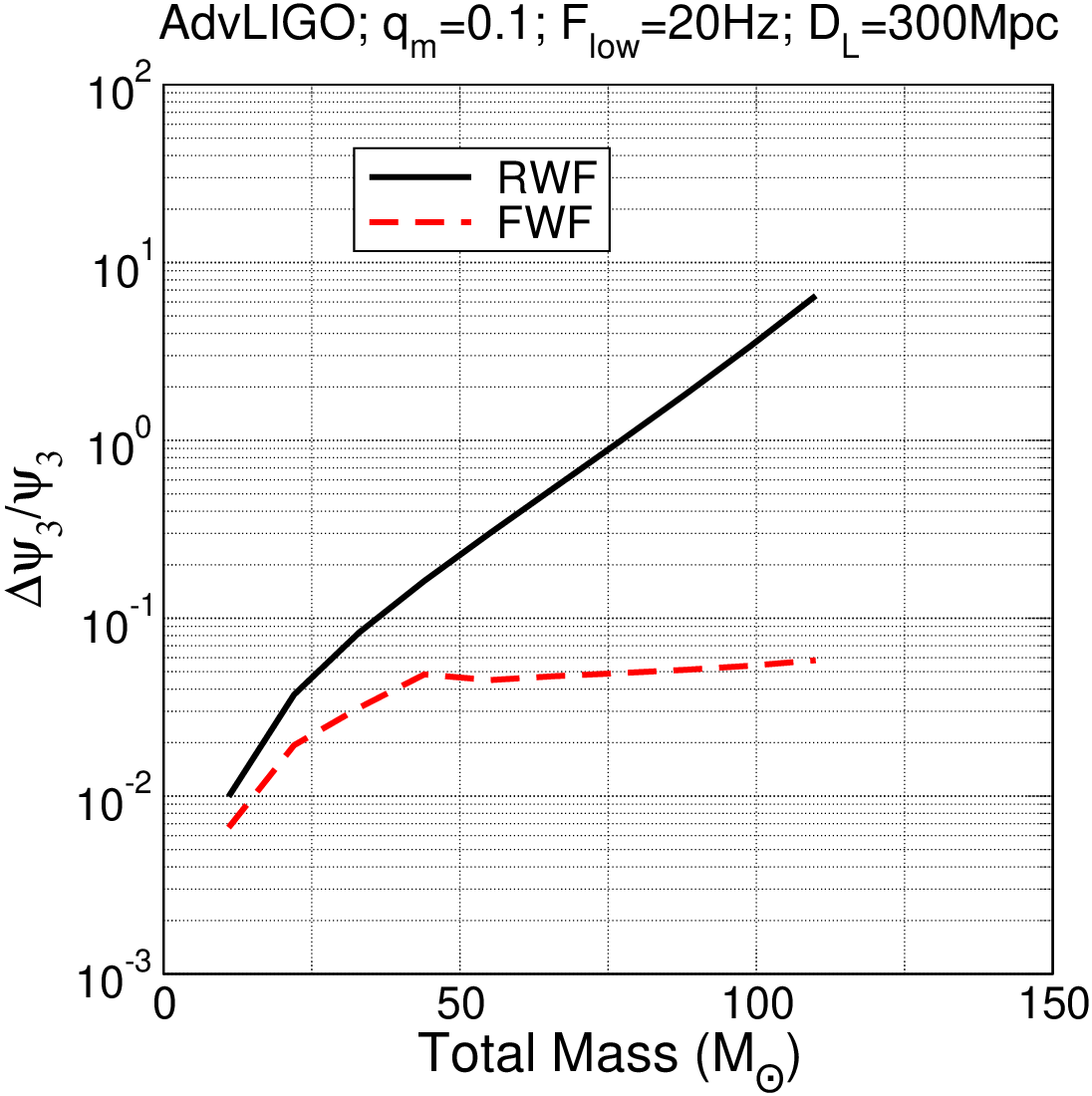}
\includegraphics[width=0.42\textwidth,angle=0]{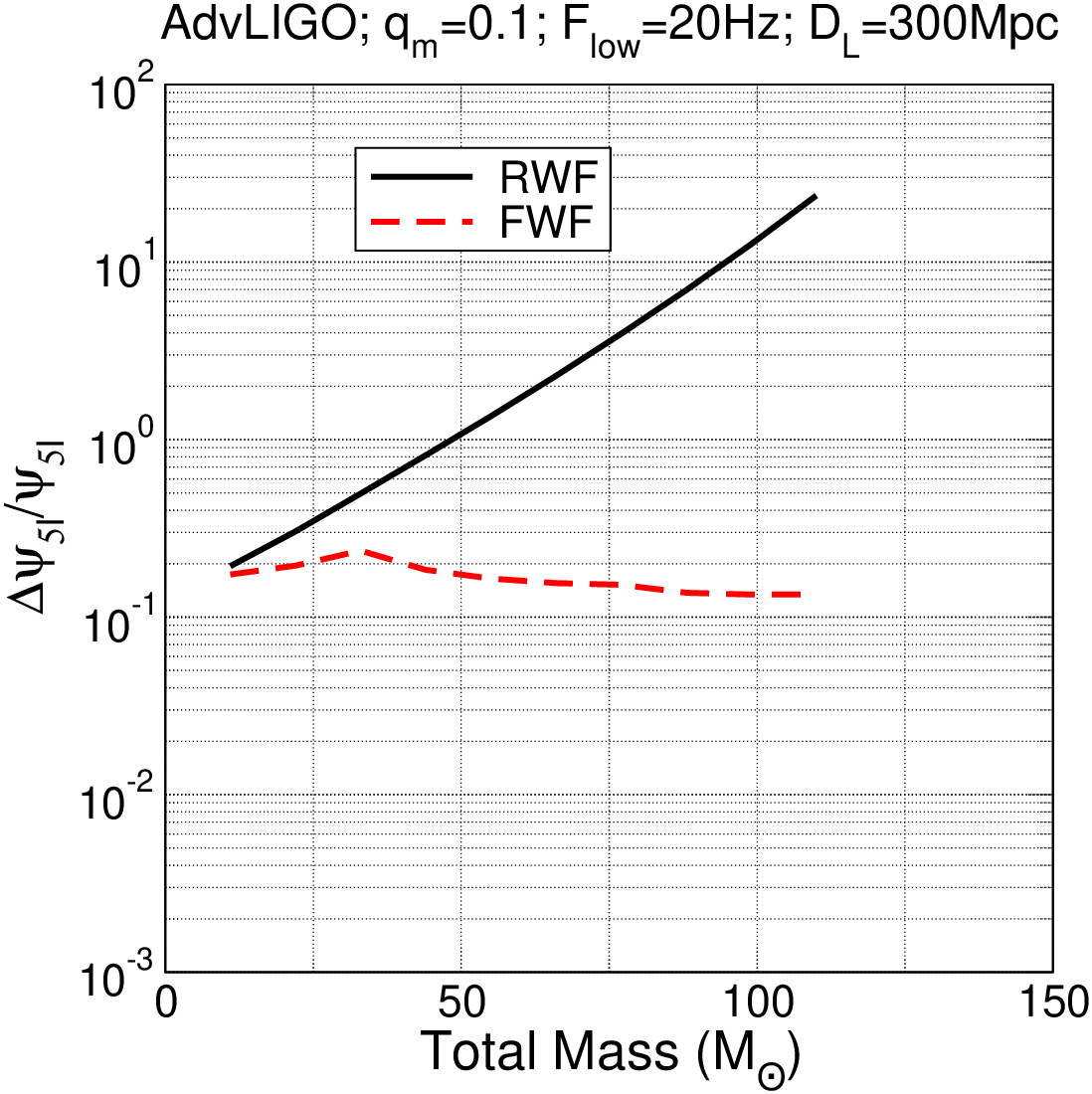}
\caption{Plots showing the variation of relative errors
$\Delta \psi_T/\psi_T$ in the test parameters $\psi_T=\psi_3, \psi_{5l}$
as a function of total mass of binaries in the range $11$-$110 M_\odot$
(with component masses having mass ratio of 0.1) located at 300 Mpc
observed by Advanced LIGO, using both the RWF and the FWF as waveform
model with the source orientations chosen arbitrarily to be
$\theta=\phi=\pi/6$, $\psi=\pi/4$, and $\iota=\pi/3$. The noise curve corresponds to the one shown in Fig.~\ref{fig:noisecurves} for the Advanced LIGO case and its analytical fit is given by Eq.\eqref{advligo-psd}. It is evident from the plot in the left panel that the fractional
accuracies with which $\psi_3$ can be measured are better than 6\%
for the entire mass range under consideration when FWF is used and
thus can be used to test the theory of gravity. $\psi_{5{\rm l}}$
(right panel) can be measured with fractional accuracies better
than 23\% for the entire mass range when FWF is used but being a
poorly determined parameter it can provide a much less stringent test of
the theory of gravity.}
\label{errvsmass_advligo}
\end{figure*}

As in the previous case (cf., Fig~\ref{demotog_psi5l}), we 
notice the departures from GR. However, these departures are 
not very clear in the middle panel where the 2.5PN coefficient 
and other higher-order terms  differ from their GR values 
by only 1\%. As a result one would not be able to discriminate 
between the two theories by using GW observations of such 
sources in ET, even though the parameter can be measured accurately.  
The right panel of the Fig.~\ref{demo_tog}, which corresponds to a theory 
in which the values of 2.5PN and higher-order coefficients differ 
from their GR values by 10\%, brings such significant departures 
from GR  that despite the larger errors in the estimation of the 
test parameter, one can distinguish between the two theories 
using the test we are proposing here. One should bear in mind that the model we have used is 
the simplest possible way in which a deviation could occur. But 
our purpose here is to have a proof of principle demonstration of 
the power of the proposed test, given the lack of knowledge of the
exact manner in which such deviations could occur. 

To summarize, assuming that GR is correct, our proposal gives the accuracy with
which three of the PN parameters can be measured. {\it How does that test GR?}
For e.g. if GR is not correct and differs at, say 1.5 PN level onwards, then
our claim is that we would begin to see inconsistencies in the estimated 
parameter values beyond the accuracy of measurement provided deviations from
GR are large enough. One may be concerned about the extent to
which the departure of higher-order terms from their GR values would 
penalize the estimation of lower-order terms. As evidenced by our examples
above, they induce bias in the estimation of parameters but \textit{do not}  
lead to greater errors in the estimation of parameters. In other words, the 
1.5PN and higher-order PN coefficients not agreeing with GR might shift the mean
of the distribution of $(M, \nu)$ but the width should remain more or less
the same.
Put  differently, if the PN expansion differs from GR slightly then
the error in the estimation of parameters will not change to first
order.

\section{The results}
\label{Results}
\subsection{Advanced LIGO}\label{results_advligo}

\begin{figure*}[t]\centering
\includegraphics[width=0.38\textwidth,angle=0]{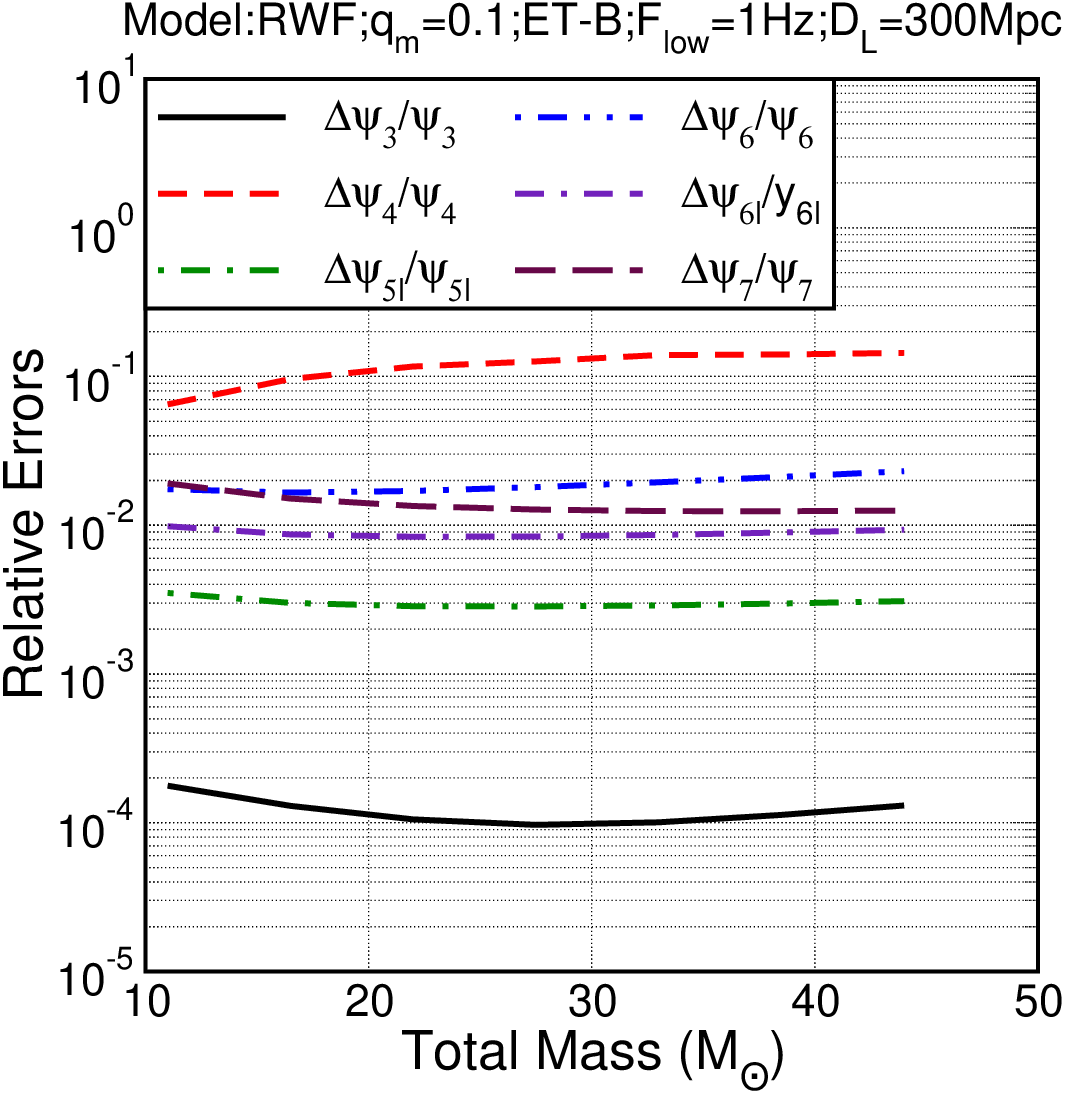}
\hskip 0.5cm
\includegraphics[width=0.38\textwidth,angle=0]{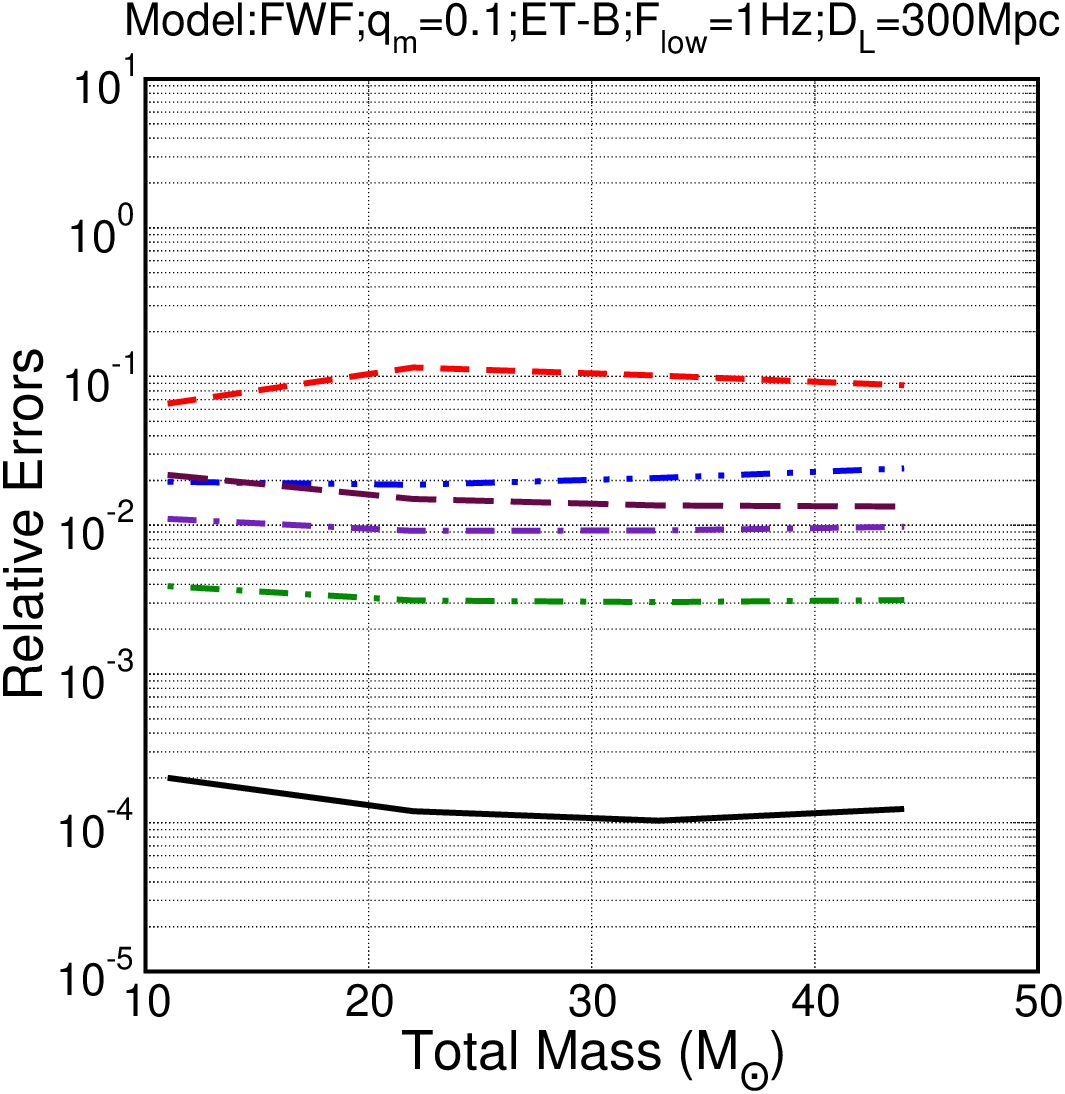}
\vskip 0.7cm
\includegraphics[width=0.38\textwidth,angle=0]{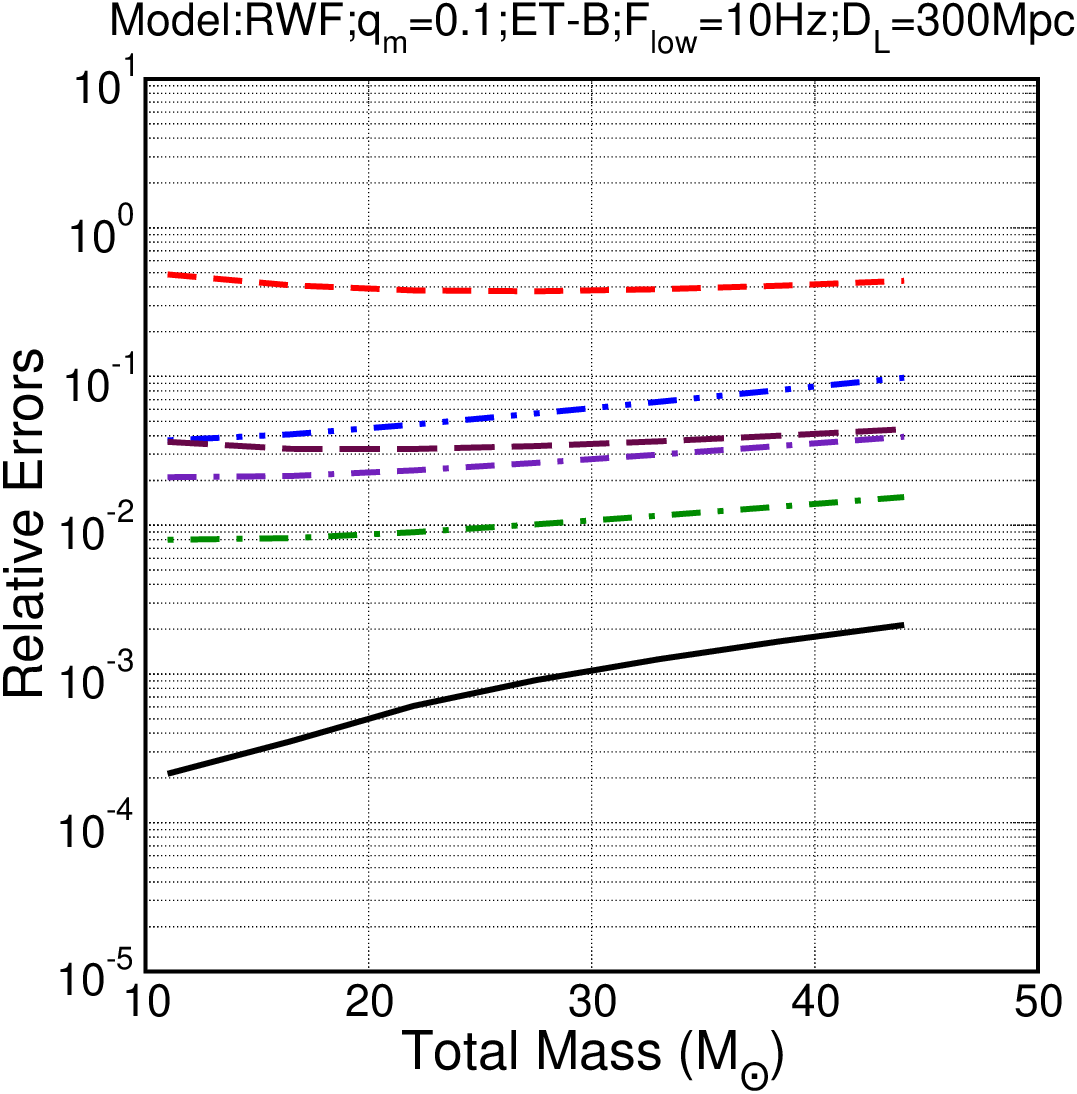}
\hskip 0.5cm
\includegraphics[width=0.38\textwidth,angle=0]{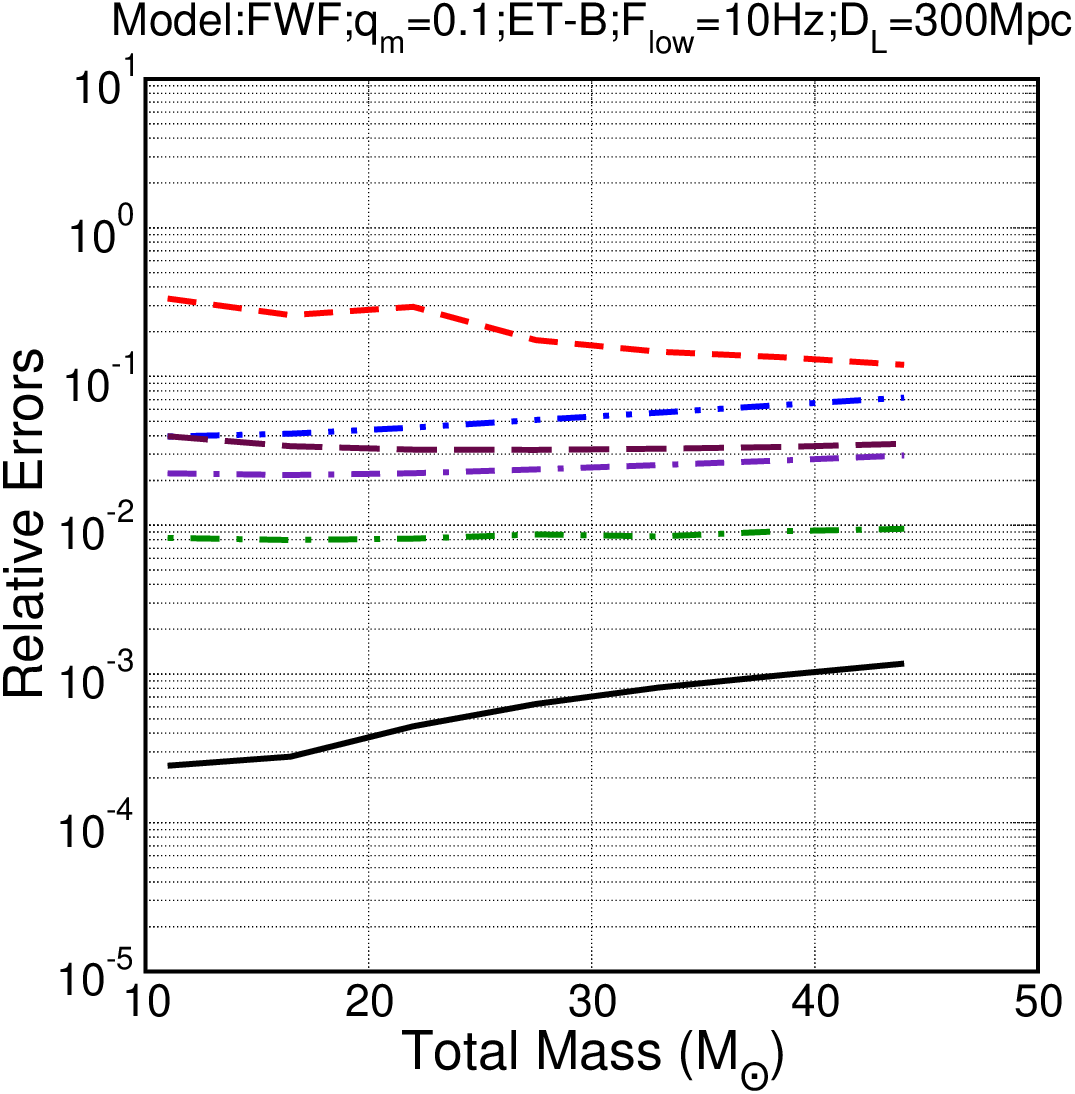}
\caption{Plots showing the variation of
relative errors $\Delta\psi_T/\psi_T$ in the test 
parameters $\psi_T$=$\psi_3$, $\psi_4$, $\psi_{5l}$, $\psi_6$, $\psi_{6l}$, and 
$\psi_7$ as a function of total mass $M$ for  stellar mass black hole 
binaries (with component masses having mass ratio 0.1) at a luminosity 
distance of $D_L=300$ Mpc observed by ET, using both RWF (left panels) 
and FWF (right panels) as  waveform models. The choice of the source 
orientations is the same as quoted in Fig.~\ref{errvsmass_advligo}. 
 The noise curve corresponds to the recent ET-B sensitivity curve. 
Top  panels correspond to the lower frequency cutoff of 1 Hz. 
By using FWF as the waveform model all $\psi_k$'s  except $\psi_{4}$  
can be tested with fractional accuracy better than 2\% in  the  mass
range $11$-$44M_{\odot}$. Bottom panels correspond to the lower 
frequency cutoff of 10 Hz. Using FWF, all $\psi_k$'s except $\psi_4$
can be tested with fractional accuracy better than 7\% in the mass range
11-$44 M_{\odot}$.} 
\label{ErrVsmass_STMBH_0p1}
\end{figure*}

 In this section we investigate the possibility of the test using 
 GW observations of BBHs in Advanced LIGO. As discussed earlier,
 the range of total mass explored is $11$-$110\,M_\odot$ 
 and we assume that binaries are located at a luminosity distance 
 of 300 Mpc. Plots in Fig.~\ref{errvsmass_advligo} show the variation 
 of relative accuracies with which two of the PN coefficients, 
 $\psi_3$ and $\psi_{5{\rm l}}$, can be measured by Advanced LIGO 
 using the restricted and the full waveform. The components of the binary have 
 the mass ratio of 0.1. 
 
 It is evident from the plots that when the 
 FWF is used, $\psi_3$ and $\psi_{5{\rm l}}$ can be measured with 
 fractional accuracies better than 6\% and 23\%, respectively, in the 
 whole mass range under consideration. We shall require (rather arbitrarily)
 that the relative error in the measurement of a PN coefficient be less 
 than 10\% in order for the test to be effective. Clearly, $\psi_3$ 
 can be estimated quite accurately and thus it can be used  to test 
 the theory. On the other hand, since $\psi_{5{\rm l}}$ is not so 
 well determined, it can still provide a less stringent
 test of the theory. 
 The measurement of other PN coefficients is not accurate enough to lead to 
 a meaningful test of GR. 
 
 The plots clearly show the benefits of bringing higher harmonics into 
 the analysis. The use of the FWF typically improves the estimation 
 by a factor of 3 to almost 100.

\subsection{Einstein Telescope}
\label{results_ET}
In the previous section we have seen that with Advanced LIGO
one can only test PN theory up to 1.5PN. Can one do better with 
the proposed third generation detector like the ET?
In what follows we investigate the extent to which one can  
test the PN theory using GW observations of stellar mass and 
intermediate mass BBHs using ET. In addition to this we will 
discuss some other key issues influencing the results such as 
effects of PN systematics on the test, choice of parametrization 
and dependence of the test on angular parameters.  

\subsubsection{Stellar mass black-hole binaries}
\label{subsec:STMBHs}

\begin{figure*}[htbp!]
\centering
\includegraphics[width=0.30\textwidth,angle=0]{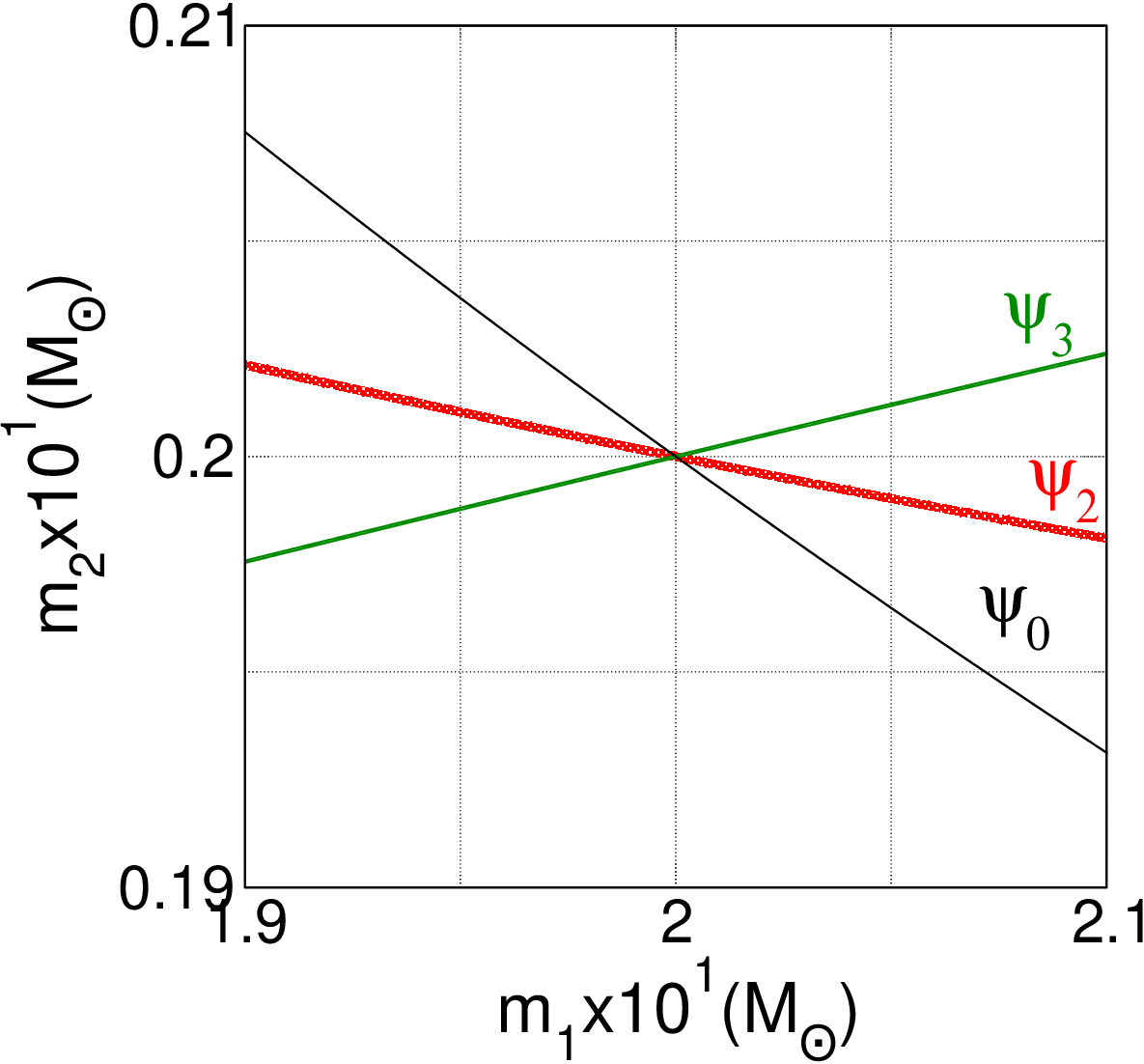}
\hskip 0.01cm
\includegraphics[width=0.32\textwidth,angle=0]{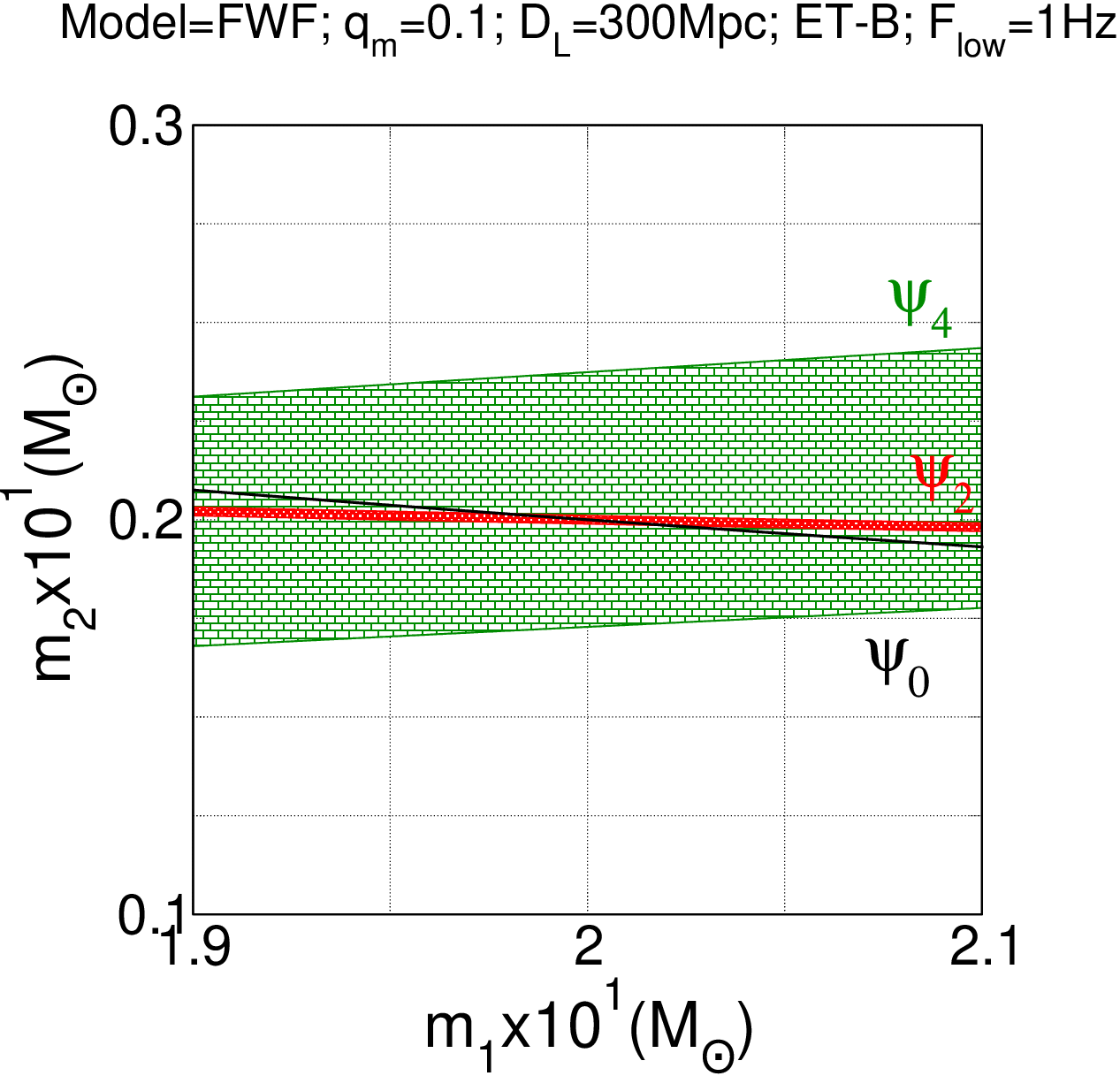}
\hskip 0.01cm
\includegraphics[width=0.30\textwidth,angle=0]{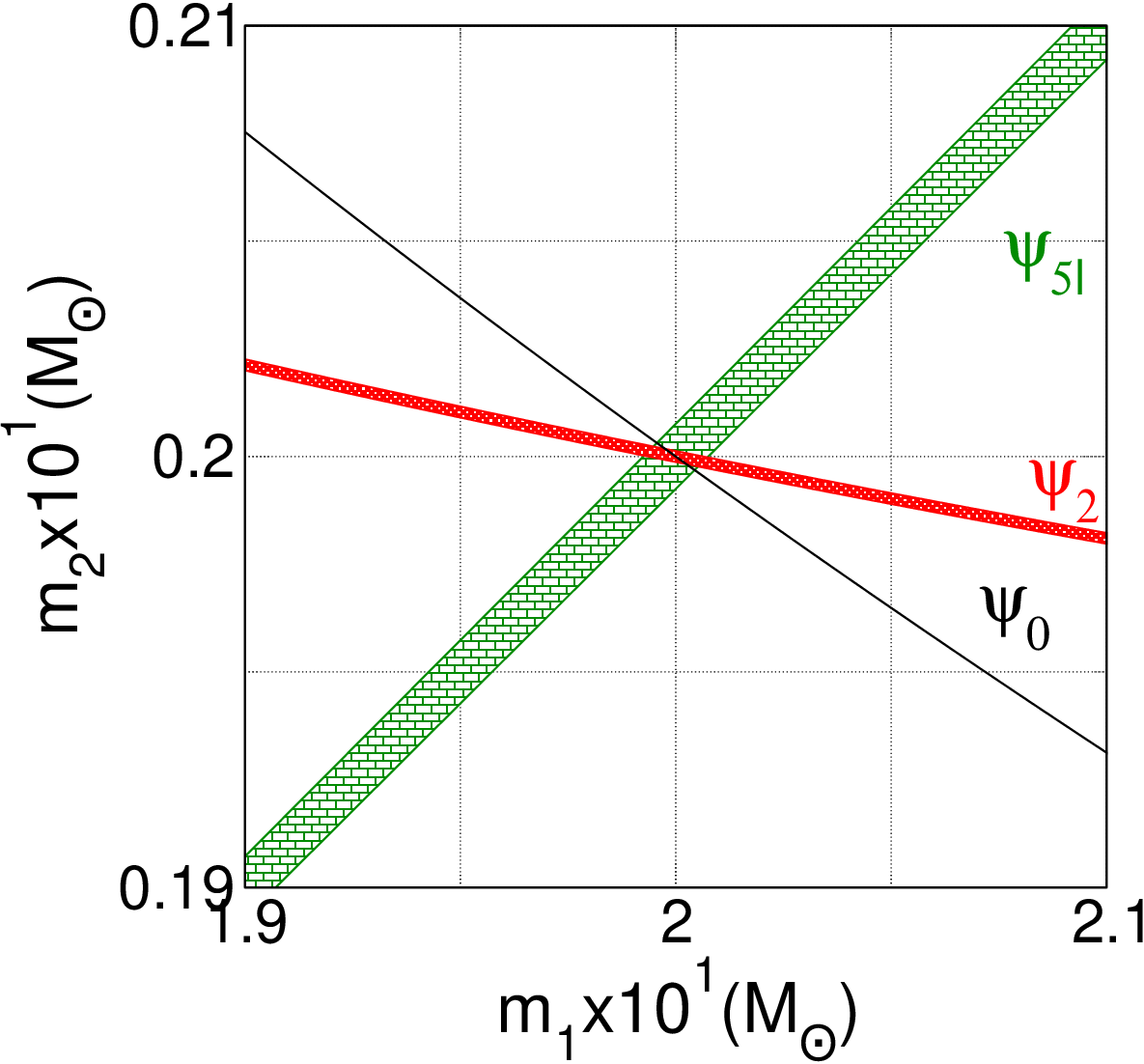}
\vskip 0.8cm
\includegraphics[width=0.30\textwidth,angle=0]{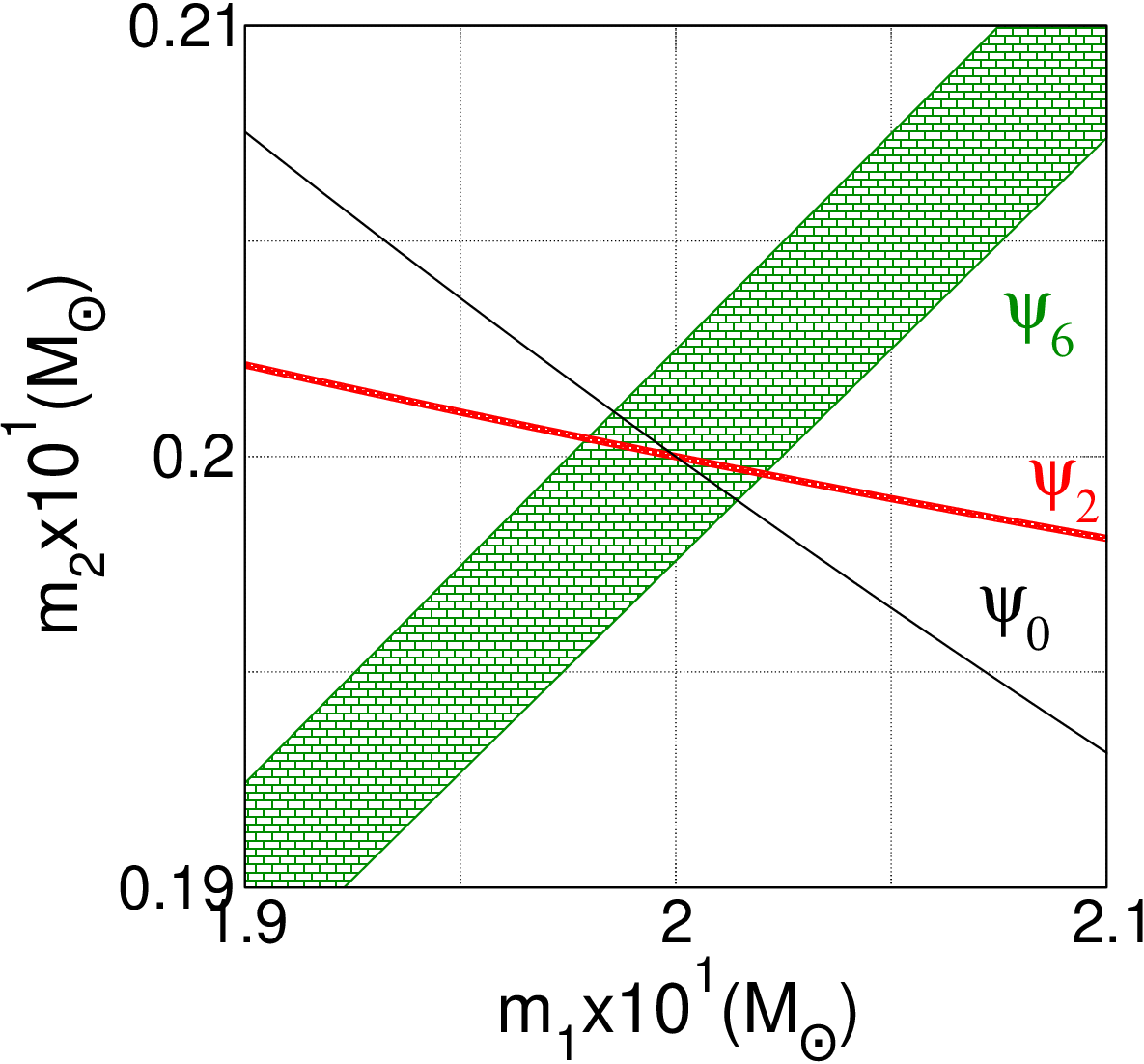}
\hskip 0.1cm
\includegraphics[width=0.30\textwidth,angle=0]{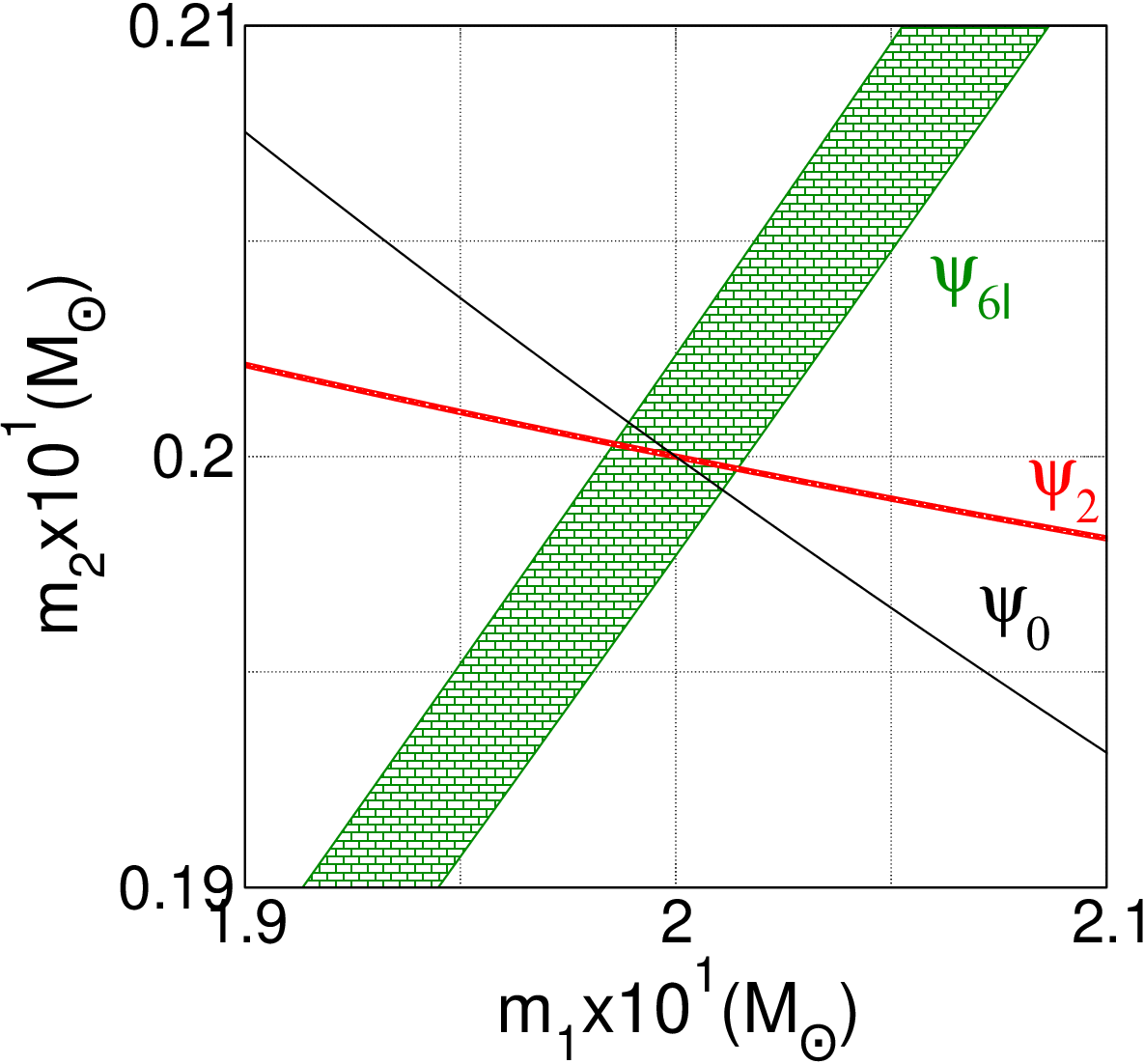}
\hskip 0.1cm
\includegraphics[width=0.30\textwidth,angle=0]{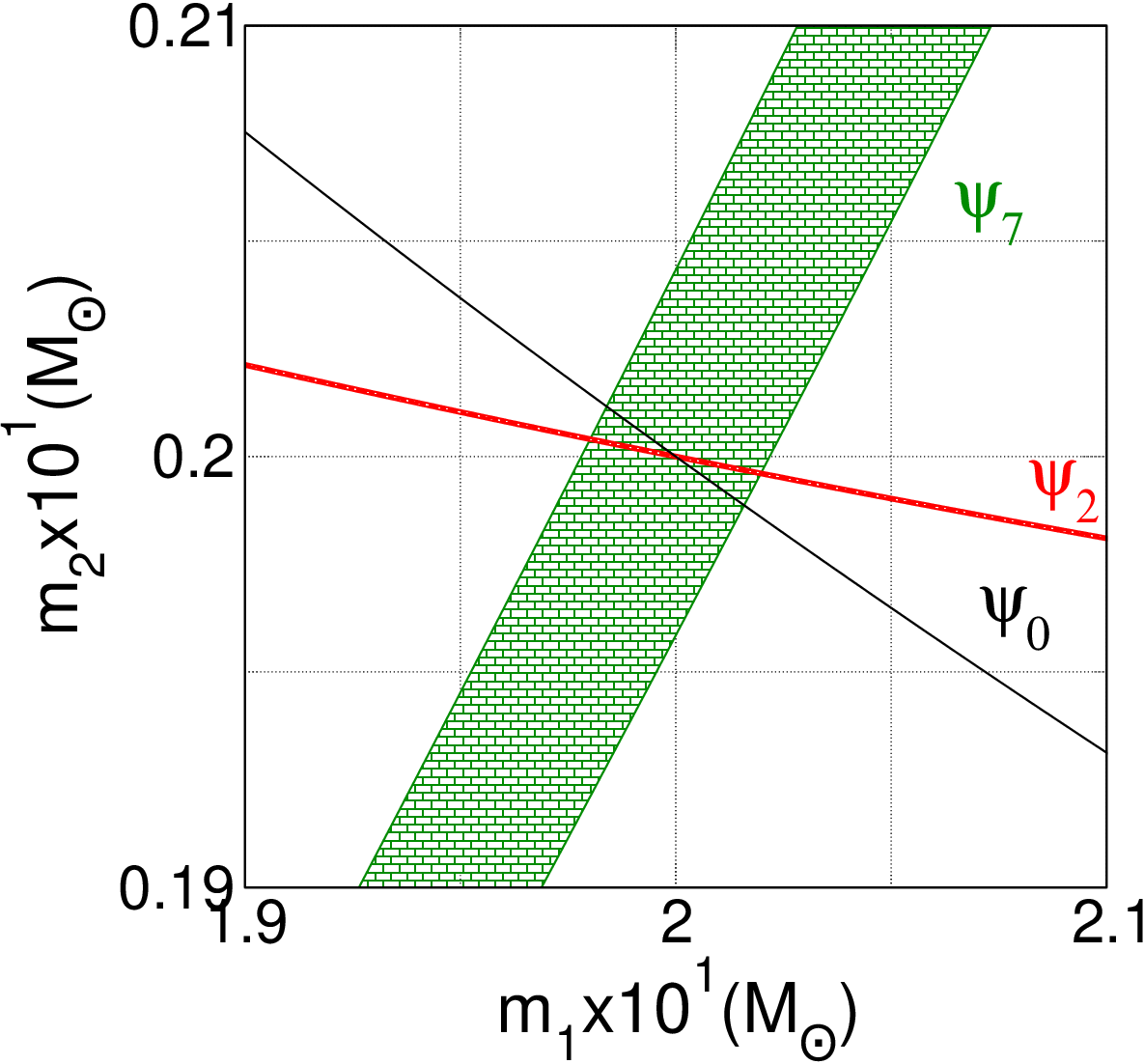}
\caption{
Plots showing the regions in the $m_1$-$m_2$ plane that correspond to
1-$\sigma$ uncertainties in $\psi_0$, $\psi_2$ and various test
parameters, which happen to be one of the six test parameters $\psi_T=\psi_3,
\psi_4, \psi_{5l}, \psi_{6}, \psi_{6l}, {\rm and}\,\psi_7$ at one time, for a (2, 20) $M_{\odot}$ BBH at
a luminosity distance of $D_L=300$ Mpc observed by ET. In all six plots shown
above  $\psi_0$ and $\psi_2$ are chosen as the  fundamental parameters (from which we can
measure the masses of the two black holes). Each parameter corresponds to a given
region in the $m_1$-$m_2$-plane 
and  if GR is the correct theory of gravity then all three parameters,
$\psi_0$, $\psi_2$ and $\psi_T$ should have a nonempty intersection in the $m_1$-$m_2$ plane. 
A smaller region leads to a stronger test.
Notice that all panels have the same scaling except the top middle panel in which 
$Y$ axis has been scaled by a factor 10.
}
\label{m1m2_22Msun}
\end{figure*}

Figure~\ref{ErrVsmass_STMBH_0p1} plots the relative errors $\Delta
\psi_T/\psi_T$ as a function of total mass $M$ of the binary at a 
distance of $D_L=300$ Mpc. We have considered stellar mass BBHs of 
unequal masses  and mass ratio 0.1, with the total mass in the range
$11$-$44 M_{\odot}$. Figure~\ref{ErrVsmass_STMBH_0p1} also 
shows two types of comparisons: (a) full waveform vs 
restricted waveform, and (b) a lower frequency cutoff of 
10 vs 1 Hz. The  top and bottom 
panels correspond to the lower frequency cutoff of 1 and 
10 Hz, respectively, while the left and right panels correspond to 
the RWF and FWF, respectively.  The source orientations are chosen 
arbitrarily to be $\theta=\phi=\pi/6$, $\psi=\pi/4$, and $\iota=\pi/3$. 
It should be evident from the plots that the best estimates of 
various test parameters are for the combination using the FWF  with 
a lower cutoff frequency of 1 Hz. In this case, all $\psi_i$'s except 
$\psi_4$ can be measured with fractional accuracies better that 
2\% for the total mass in the range $11$-$44M_{\odot}$. On 
the other hand when the lower cutoff is 10 Hz, with the 
FWF all $\psi_i$'s except $\psi_4$ can be measured with fractional 
accuracies better than 7\%. It is also evident from the plots
that as compared to other test parameters, $\psi_3$ is the most 
accurately measured parameter in all cases and best estimated when 
the lower frequency cutoff is 1 Hz.  On the other hand, $\psi_4$ is the 
worst measured parameter of all the test parameters.
However, we see the best improvement in its measurement when 
going from the RWF to the FWF.

Figure~\ref{m1m2_22Msun} shows the regions in the $m_1$-$m_2$ plane that
correspond to 1-$\sigma$ uncertainties in $\psi_0$, $\psi_2$ and 
various test parameters which in turn will be one of the six test 
parameters $\psi_T=\psi_3, \psi_4, \psi_{5l}, \psi_{6}, \psi_{6l}, {\rm and}\,\psi_7,$ one at a time, for a (2, 20) $M_{\odot}$ BBH, at a luminosity 
distance of $D_L=300$ Mpc observed by ET. It is evident from the plots
corresponding to various tests that each test parameter is consistent 
with corresponding fundamental pair ($\psi_0$, $\psi_2$).\\
\begin{figure*}[htbp!]
\centering
\includegraphics[width=0.38\textwidth,angle=0]{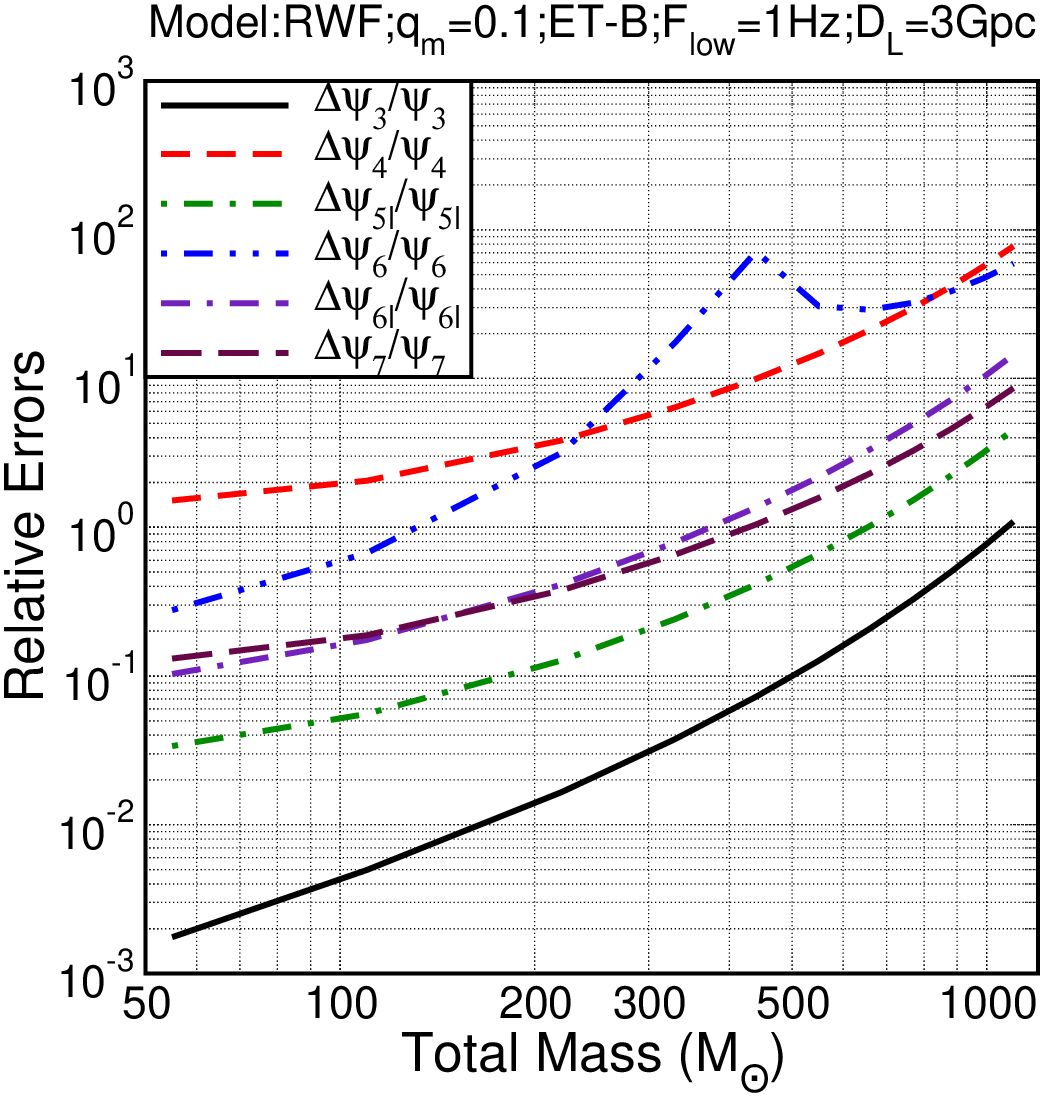}
\hskip 0.5cm
\includegraphics[width=0.38\textwidth,angle=0]{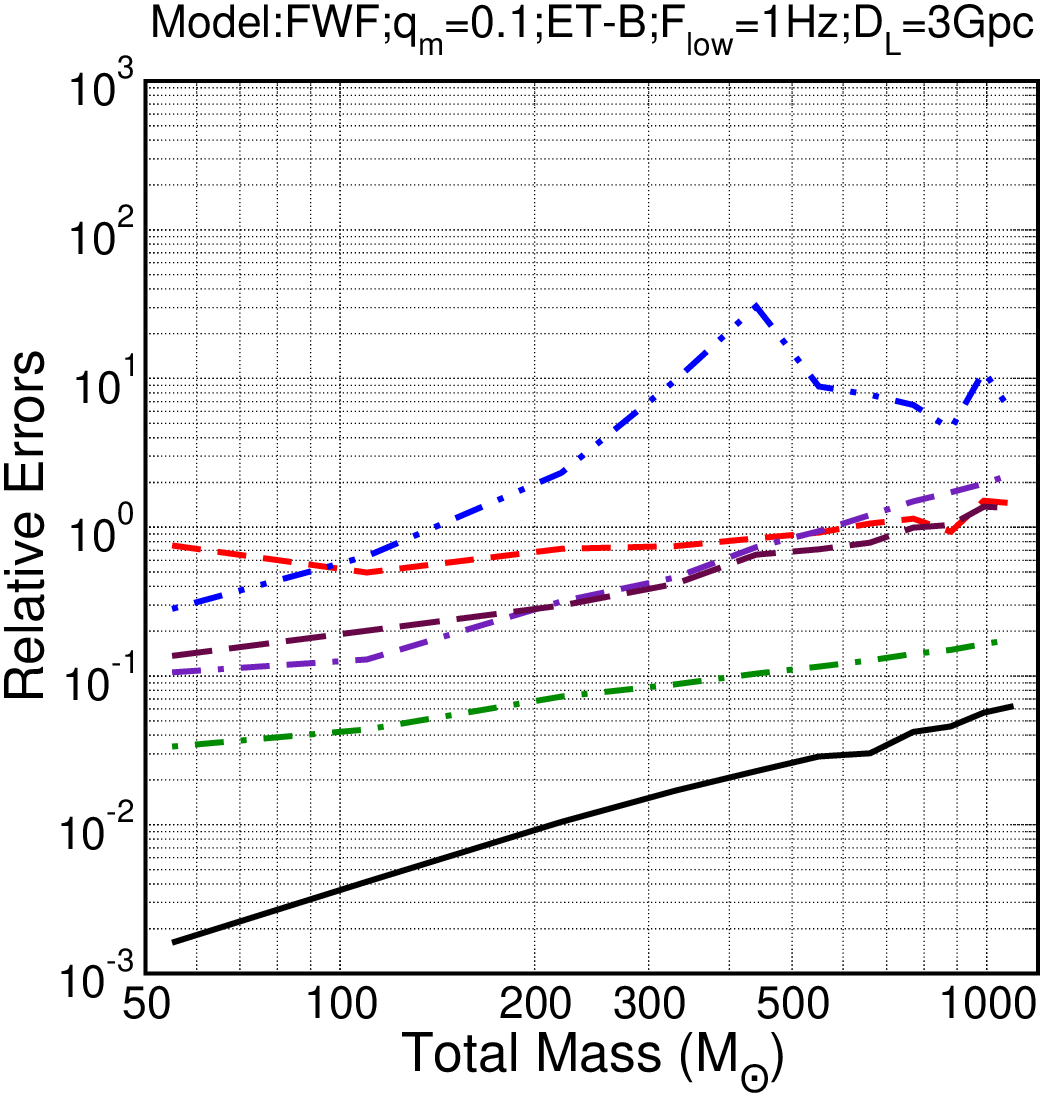}
\vskip 0.7cm
\includegraphics[width=0.38\textwidth,angle=0]{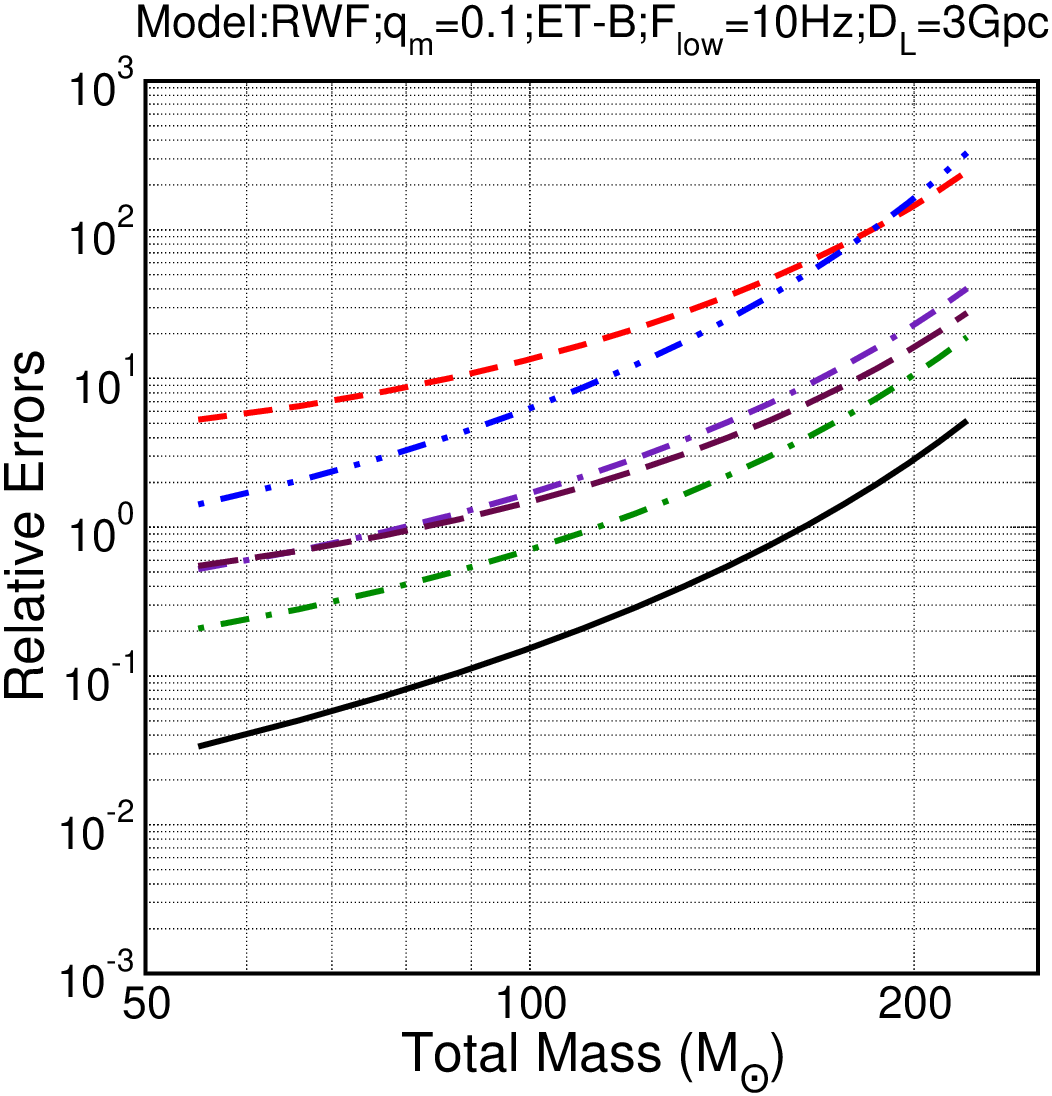}
\hskip 0.7cm
\includegraphics[width=0.39\textwidth,angle=0]{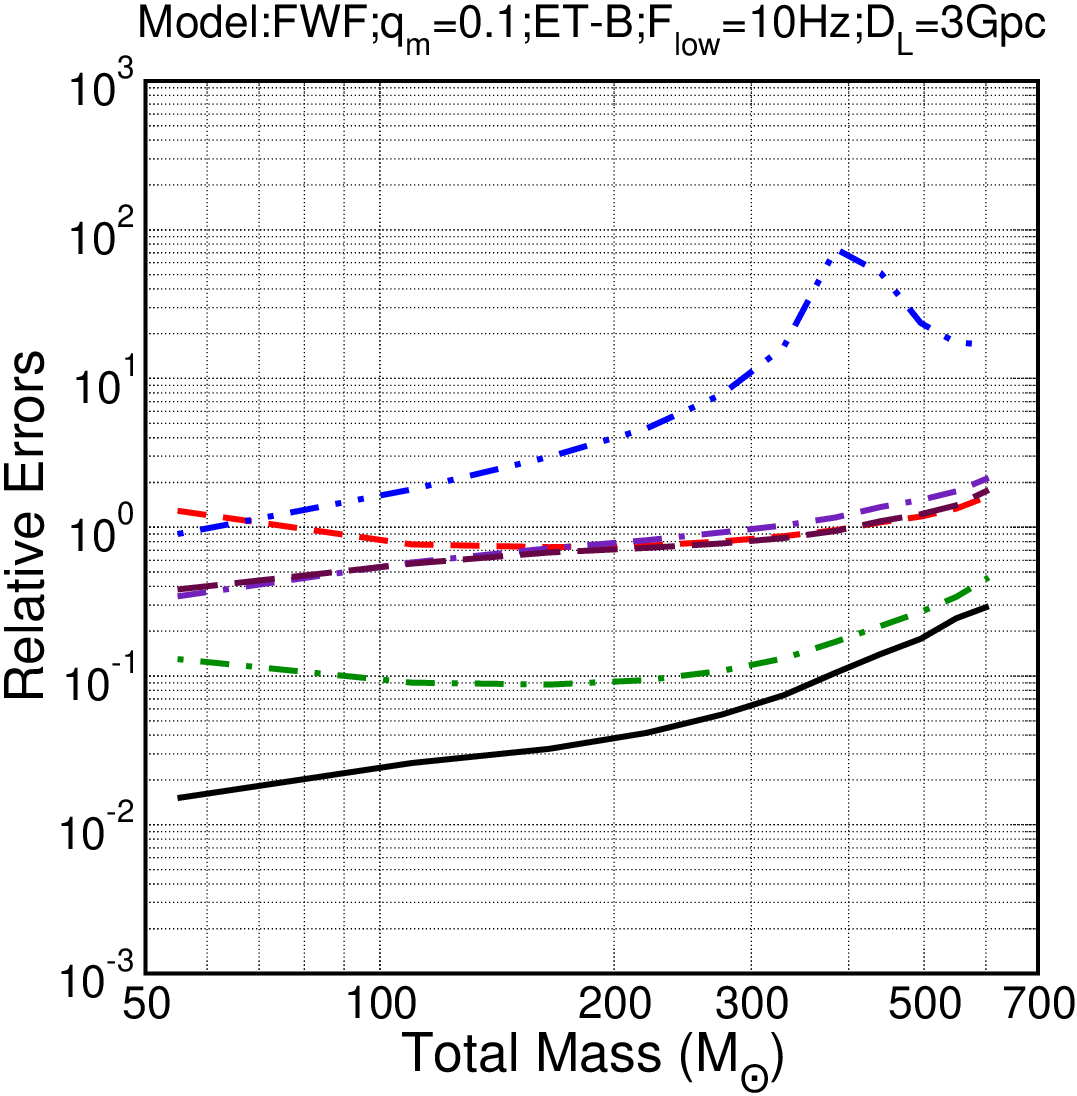}
\caption{Same as Fig.\ref{ErrVsmass_STMBH_0p1} but for 
intermediate mass black hole binaries (with component masses having mass
ratio 0.1) at a luminosity distance of $D_L$=3 Gpc. With lower frequency 
cutoff of 1 Hz, using FWF as the waveform model, $\psi_{3}$ and $\psi_{5l}$  
can be tested with fractional accuracy better than 10\% for the mass
range 55-$400 M_{\odot}$. On the other hand, with a lower frequency cutoff 
of 10 Hz, using the  FWF, $\psi_3$ and $\psi_{5l}$ can be tested with 
fractional accuracy better than 10\% for the mass range
90-220$M_{\odot}$. 
}
\label{ErrVsmass_IMBH_0p1}
\end{figure*}

\subsubsection{Intermediate mass black hole binaries}
\label{IMBH}
Figure~\ref{ErrVsmass_IMBH_0p1} plots the relative errors $\Delta \psi_T/\psi_T$
as a function of the total mass $M$ of the binary at a distance of
$D_L$=3\,Gpc. We have considered BBH of unequal masses with mass
ratio 0.1. As in Fig.~\ref{ErrVsmass_STMBH_0p1},
Fig.~\ref{ErrVsmass_IMBH_0p1} also shows two types of comparisons: (a) the effect
of the use of FWF on parameter estimation against RWF, (b) the effect of lowering
the cutoff frequency from 10 to 1 Hz. As before, top and bottom panels
correspond to the cutoff frequency of 1 and 10 Hz, respectively, and left
and right panels to RWF and FWF, respectively.  The source orientations  
are chosen arbitrarily to be $\theta=\phi=\pi/6$, $\psi=\pi/4$, and $\iota=\pi/3$. 

It is evident from the plots that the least relative errors in various 
test parameters are for the combination that uses the FWF and a lower 
cutoff of 1 Hz. Unlike the case of stellar mass BBHs, in the
case of intermediate mass BBHs only two of the test parameters, $\psi_3$ and
$\psi_{5l}$, can be measured with fractional accuracies better that 10\% for
the total mass in the range 55-400 $M_{\odot}$ with FWF and lower
cutoff frequency as 1 Hz. On the other hand, when the lower frequency cutoff is
10 Hz the use of the FWF allows the estimation of $\psi_3$ and $\psi_{5l}$ with
fractional accuracies better than 10\% for the total mass in the range
90-220 $M_{\odot}.$ As compared to other test parameters, $\psi_3$ is the 
most accurately measured parameter in all cases and best estimated when 
the low-frequency cutoff is 1 Hz.  Parameters $\psi_4$ and $\psi_6$ are poorly 
measured as compared to the other test parameters but again we
see the  best improvement in the estimate of $\psi_4$ when using
the FWF.
 
\begin{figure*}[htbp!]
\includegraphics[width=0.30\textwidth,angle=0]{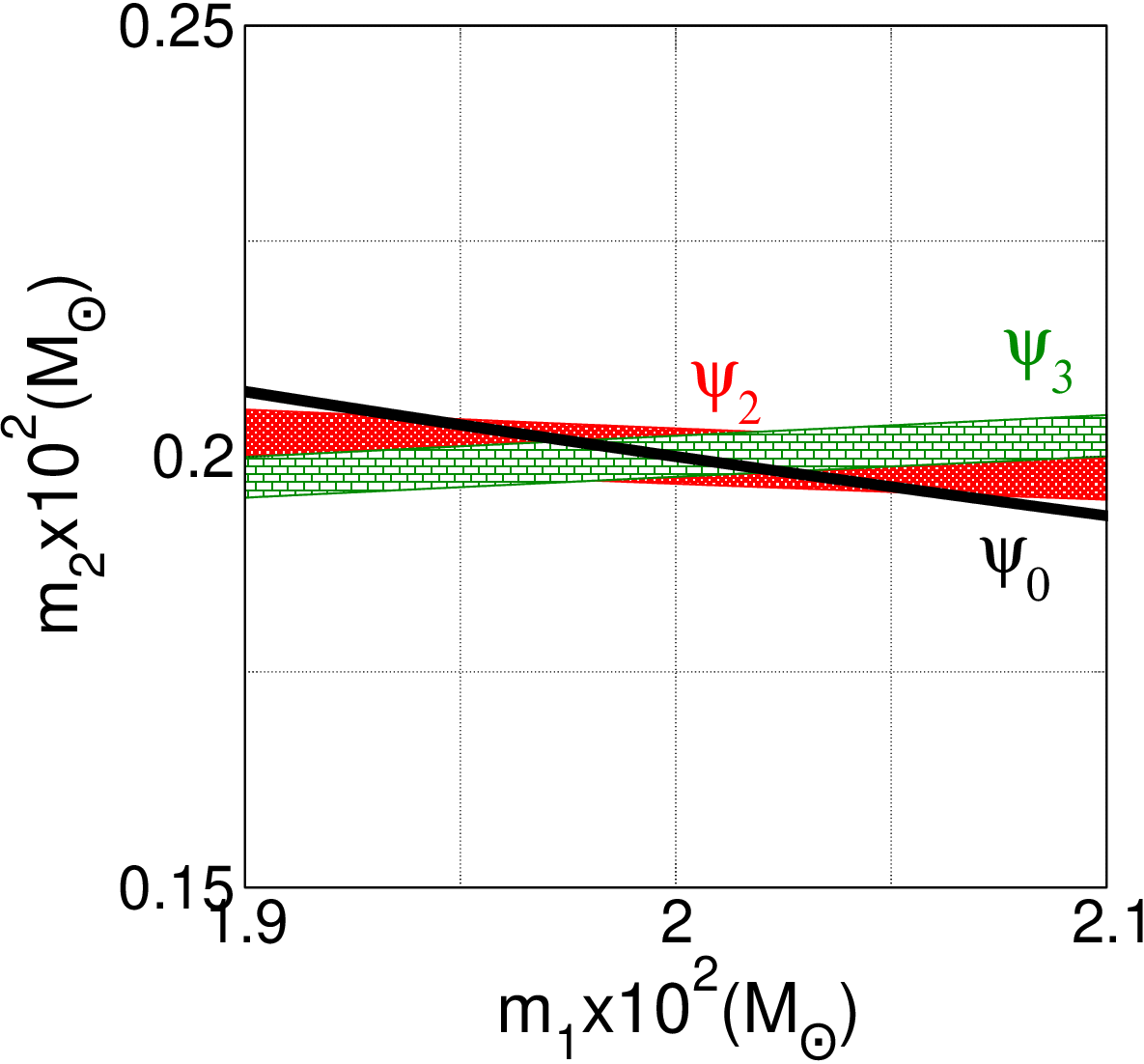}
\hskip 0.01cm
\includegraphics[width=0.32\textwidth,angle=0]{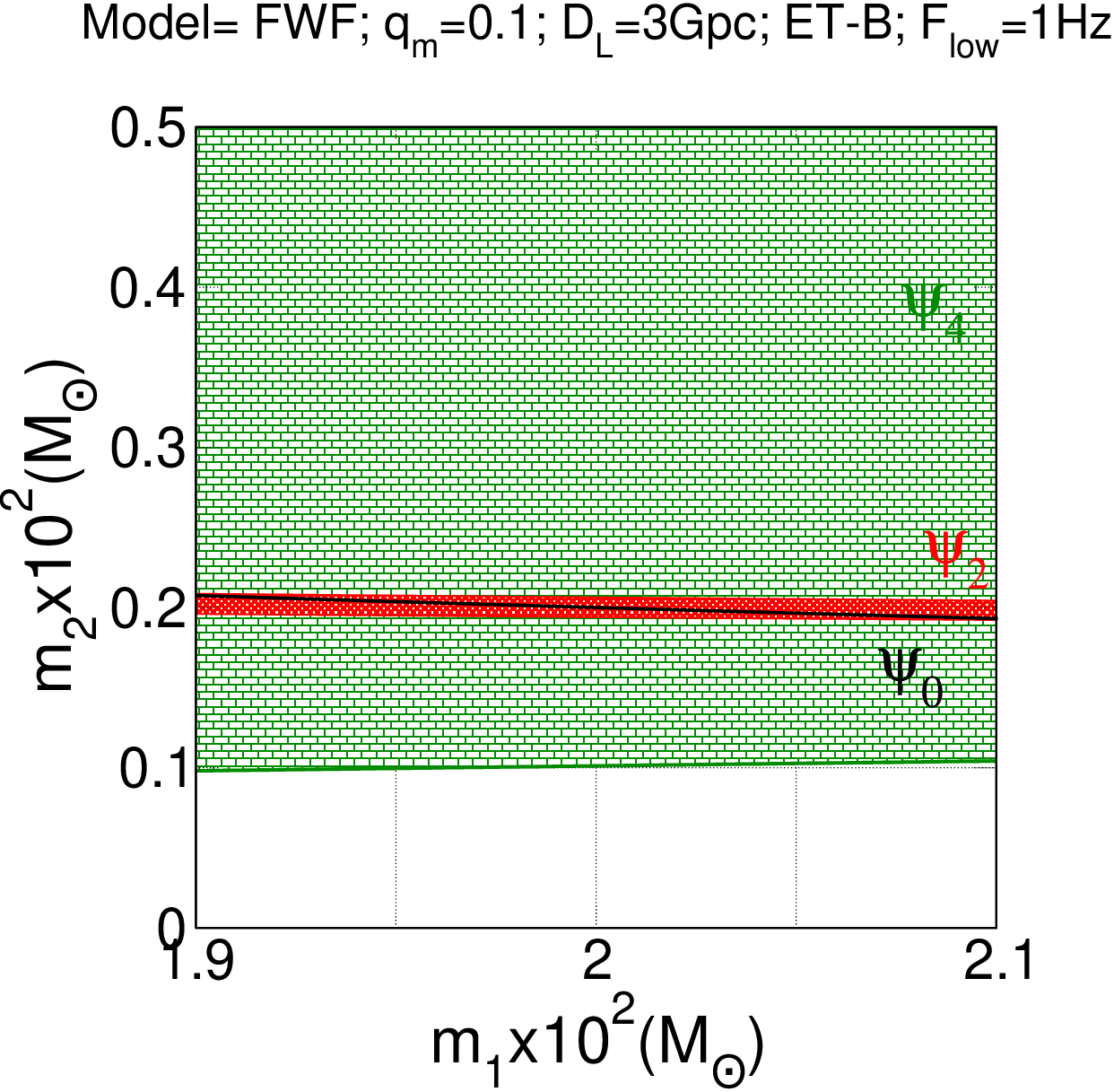}
\hskip 0.01cm
\includegraphics[width=0.30\textwidth,angle=0]{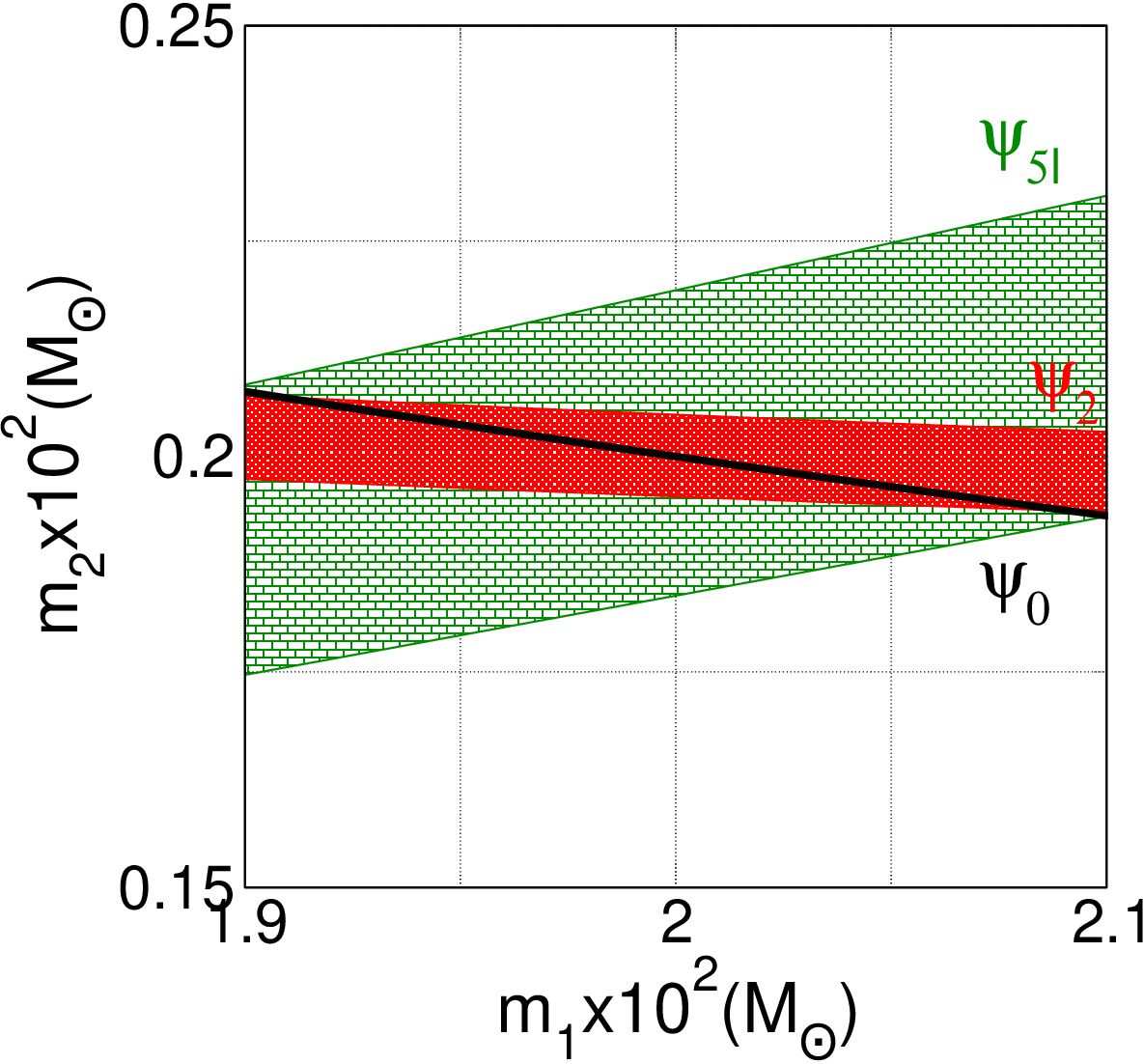}
\vskip 0.8cm
\includegraphics[width=0.30\textwidth,angle=0]{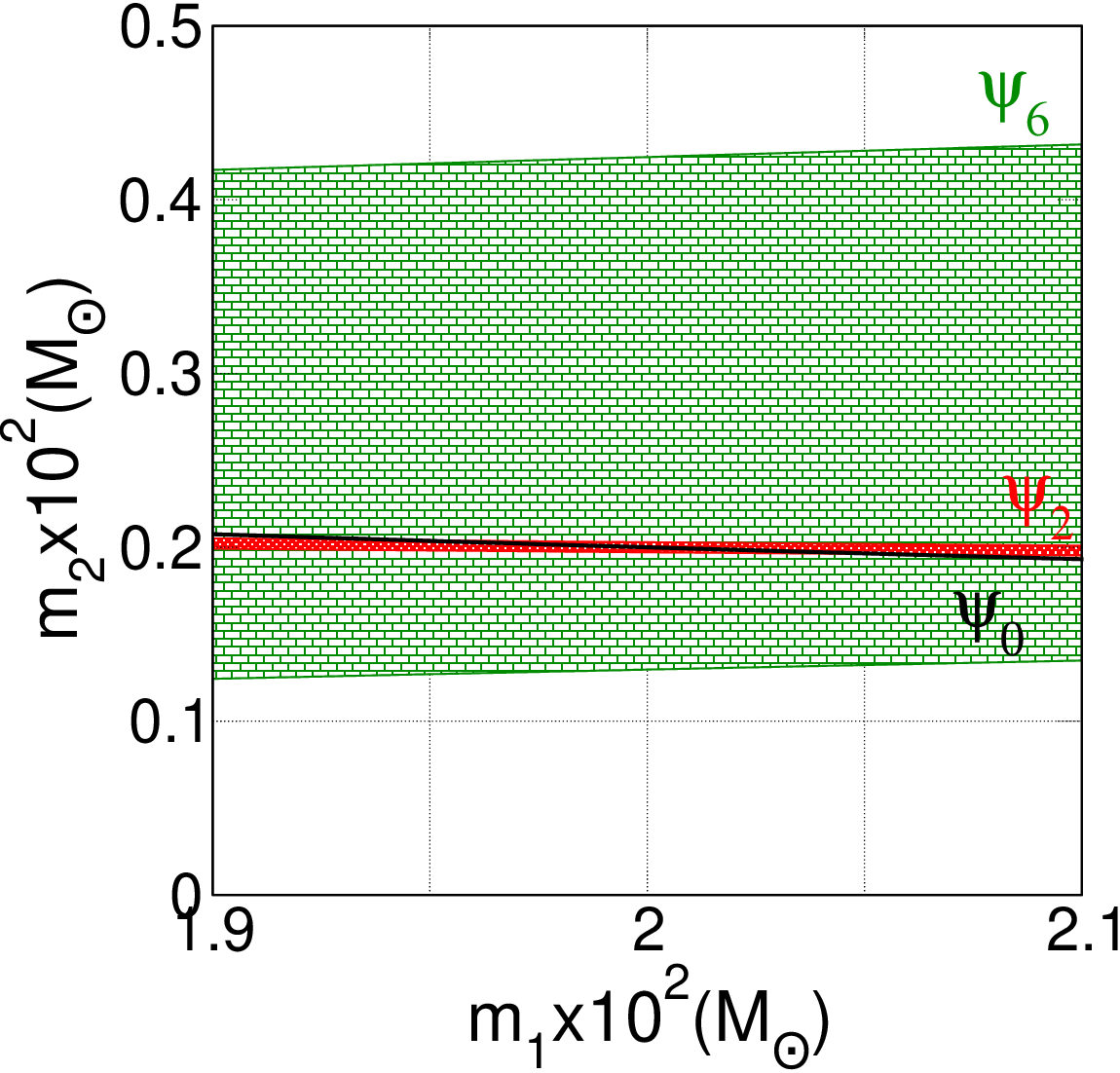}
\hskip 0.1cm
\includegraphics[width=0.30\textwidth,angle=0]{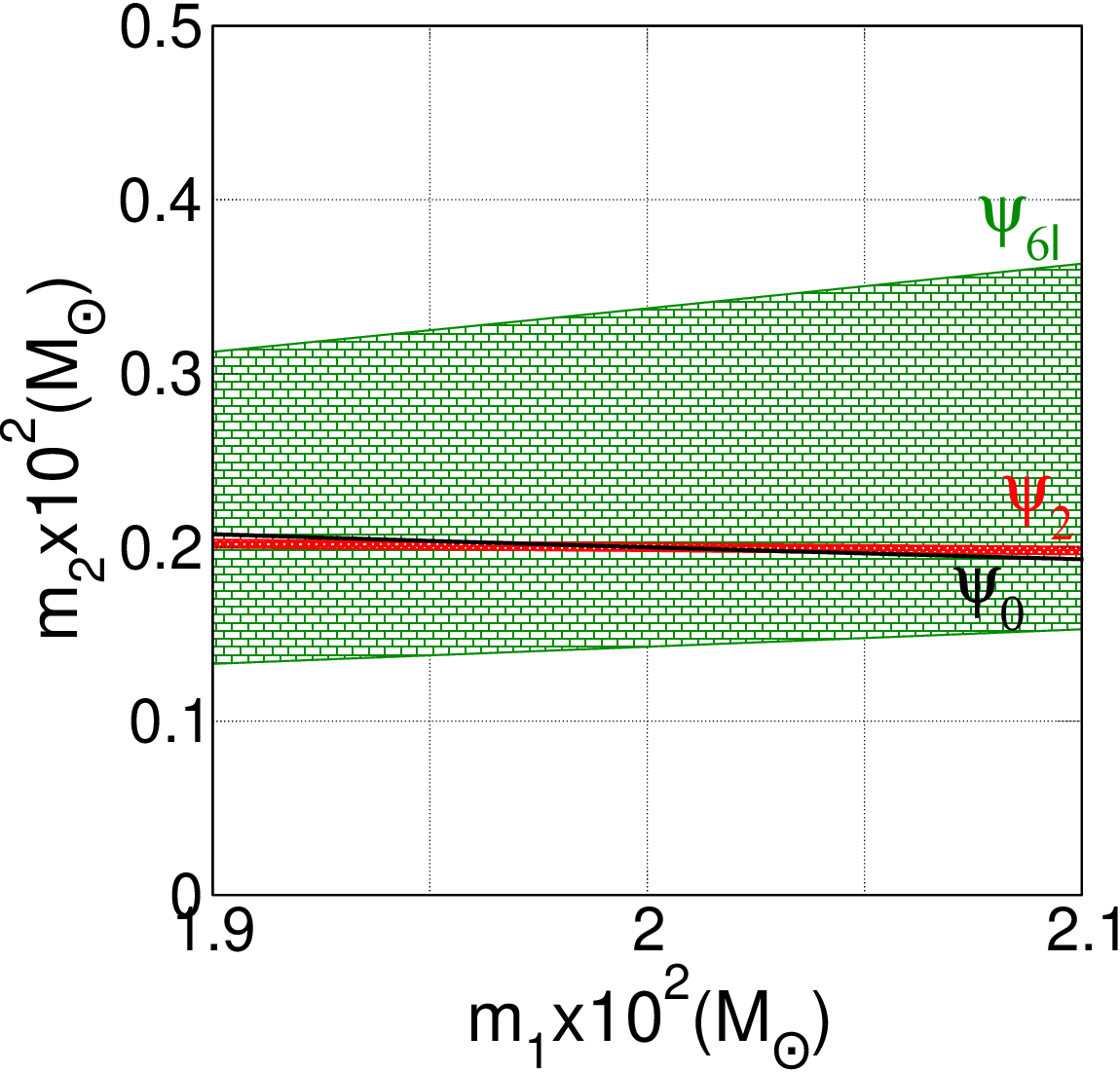}
\hskip 0.1cm
\includegraphics[width=0.30\textwidth,angle=0]{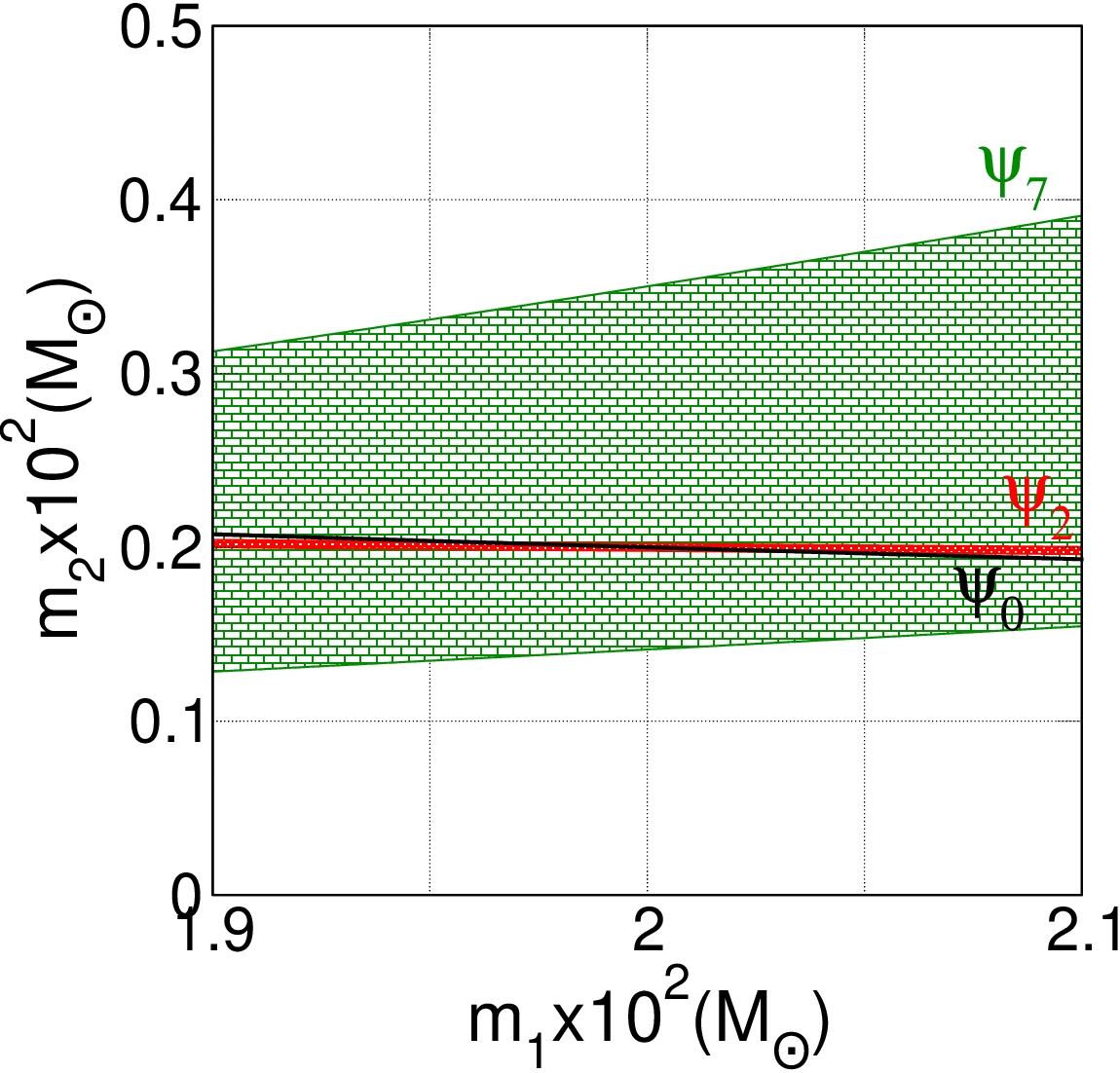}
\caption{Same as in Fig. \ref{m1m2_22Msun} but for intermediate black hole 
binaries in the mass range (20, 200)$M_{\odot}$ at a luminosity distance of 
$D_L$=3\,Gpc observed by ET. 
Notice that all bottom panels and the top middle panel have the same scaling
whereas the $Y$ axes of the top left panel and the top right panel have been scaled by a 
factor of 5. Note that there appears just one boundary for $\psi_4$ in the 
plot shown in top middle panel since the other bound does not exist for the 
range of values on $X$ axis.
}
\label{m1m2_220Msun}
\end{figure*}

Figure~\ref{m1m2_220Msun} shows the regions in the $m_1$-$m_2$ plane that
correspond to 1-$\sigma$ uncertainties in $\psi_0$, $\psi_2$ and the test
parameters $\psi_T=\psi_3, \psi_4, \psi_{5l}, \psi_{6}, \psi_{6l}, \psi_7,$  
one at a time, for a (20, 200) $M_{\odot}$ BBH at a luminosity distance of 
$D_L$=3\,Gpc observed by ET. It is clear from the plots that each test 
parameter is consistent with the corresponding fundamental pair 
($\psi_0$, $\psi_2$).

\subsubsection{Effects of PN systematics on the test}\label{sec:PNsys}
The inability to measure all the PN parameters simultaneously led us to
propose a more modest procedure to test the PN parameters one at a time. 
In parameter estimation, it  seems intuitive not to ignore 
our knowledge of the known high PN order phasing.  Further, it is natural 
to assume that if an alternative theory of gravitation, similar to GR, 
agrees with GR at some PN order, it would agree with it at a {\it lower} 
PN order but may differ from it at some {\it higher} PN order.  Thus, 
when testing a coefficient at some particular PN order, expressing a 
lower-order PN coefficient in terms of the basic pair of PN variables 
seems reasonable. However, expressing the higher PN order coefficients 
in terms of the basic pair may appear more disconcerting. 
Here we look at the issue in a little more detail and provide 
our point of view.  

We propose a  comparison of the following two schemes: The first 
scheme, as before, uses $\psi_0$ and $\psi_2$ as basic parameters 
and {\it all} known PN coefficients up to 3.5PN, except the 
test parameter, are expressed in terms of $\psi_0$ and $\psi_2.$  
The second scheme is similar but the phase evolution is truncated 
at the PN order corresponding to the test parameter.

\begin{figure*}[htbp!]
\includegraphics[width=0.65\textwidth,angle=0]{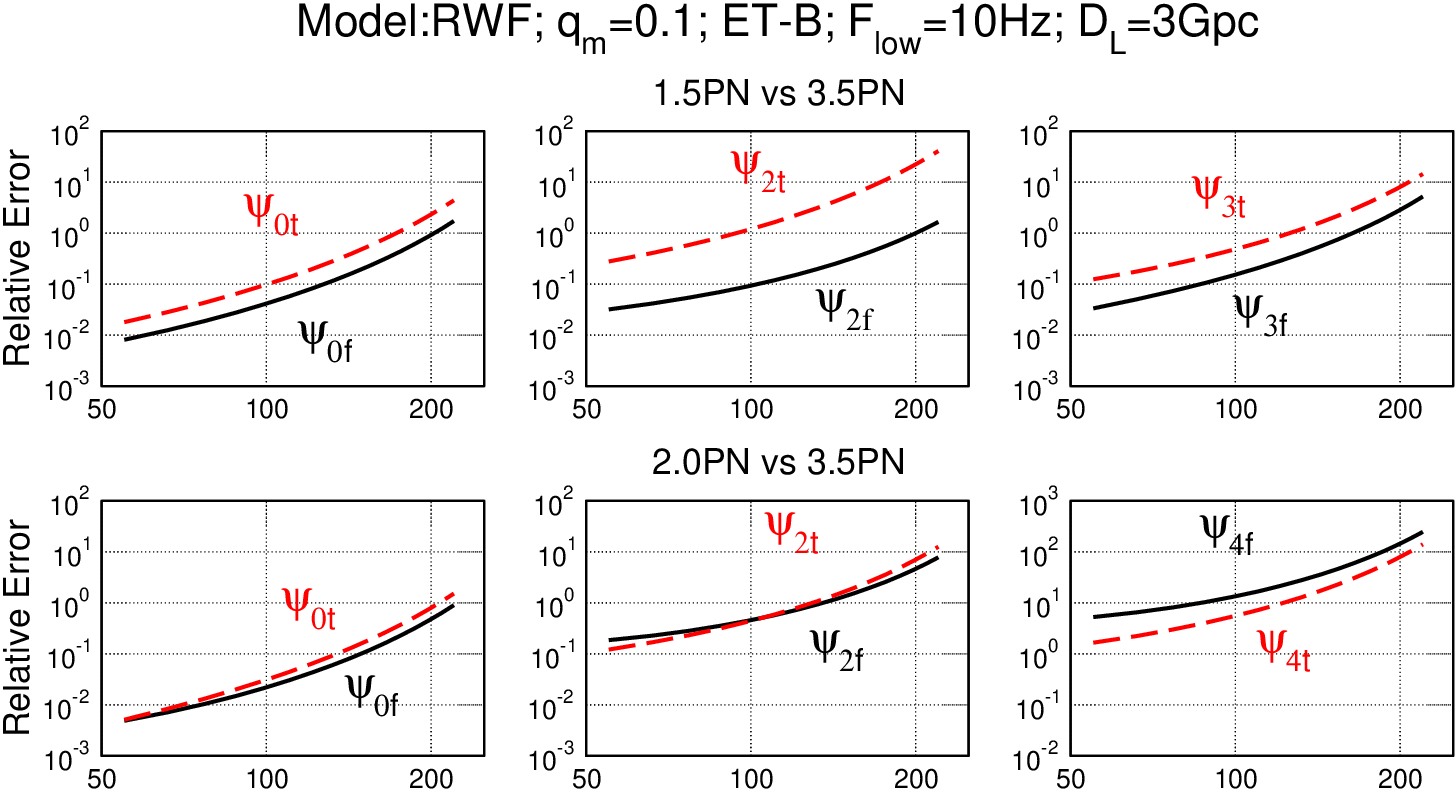}
\vskip 0.8cm
\includegraphics[width=0.65\textwidth,angle=0]{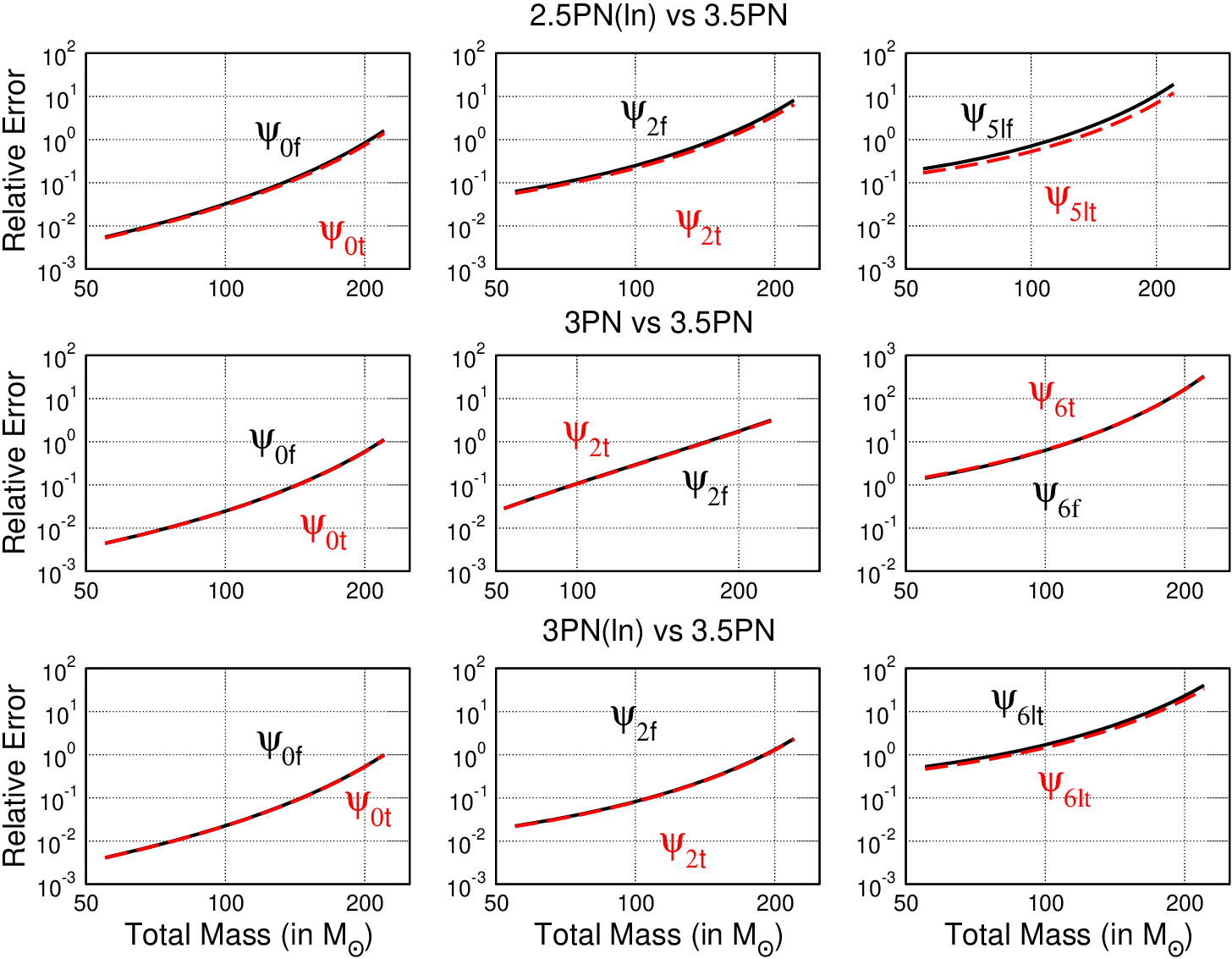}
\caption{A comparison of relative errors in the measurement of  various 
PN parameters for binaries with masses in the range 55-220$M_{\odot}$ 
at a luminosity distance of 3 Gpc for two cases: The first is the same as before when 
$\psi_0$ and $\psi_2$ are basic parameters and all PN parameters up to 3.5PN 
(full phasing) except the test parameter are parametrized by $\psi_0$ and $\psi_2$.
The others are similarly constructed but the phasing is truncated at the PN order 
corresponding to the test parameter. The low-frequency cutoff is 10 Hz and  
the RWF has been used.  The test parameter with truncated phasing is denoted 
by $\psi_{i {\rm t}}$ while with full 3.5PN phasing it is denoted by $\psi_{i{\rm f}}$.}
\label{trncvsfull}
\end{figure*}

Thus, to test $\psi_3$ in the second scheme the phasing is truncated at 
1.5PN, to test $\psi_4$ at 2PN and so on. Figure~\ref{trncvsfull} 
compares the two schemes.  It should be evident from the figure 
that in the first scheme the use of 3.5PN phasing, rather than a 
lower PN order ({ e.g.}, 1.5PN in testing $\psi_3$), does improve 
the accuracy with which one can measure a certain parameter. Conversely,
the poorer estimate ({ i.e.} larger error) in the second scheme is 
due to the neglect of higher PN order terms.

One can infer, therefore, that what is achieved in the first scheme is an 
improvement in parameter estimation arising from higher-order PN phasing;
parametrization of higher-order PN coefficients in terms the two basic 
variables is indeed a reasonable choice if one wants to quantitatively 
look at the deviation from standard GR of an alternative theory of 
gravitation similar to GR\footnote{A variant of the test starting from GW 
phasing expressions in a more general {\it meta theory} may need to be 
implemented to include theories with  different PN structure like 
scalar-tensor theories with qualitatively different effects like 
dipole radiation.}. Let us also note that if a theory of gravity deviates
from GR at a particular PN order, then in our first scheme the test
may actually fail at a lower PN order. As a result, our test will not
be able to conclusively assert the PN order that is inconsistent with GR.
Rather a failure of our test is indicative of the failure of GR at some
PN order. It would then be necessary to carry out a more powerful test 
of the theory by treating all the PN coefficients as independent parameters.
Such a test could in principle help determine at which order(s) the
true theory of gravity is inconsistent with GR.

\subsubsection{The choice of basic parametrization and the accuracy of the test}

As discussed in Sec.~\ref{Introduction}, in the present work we have
chosen the lowest-order (and hence the best determined) PN coefficients $\psi_0$ 
and $\psi_2$ to parametrize the waveform.  One might  wonder  whether 
the choice of $\psi_0$ and $\psi_2$ as basic variables is the most optimal.
To investigate this further, we explored  other choices of the basic pair
to parametrize the waveform, { e.g.} $(\psi_0,\psi_f)$ where $f$ can be
one of $(3,\,4,5l,\,6,\,6l,\,7)$.
Table~\ref{Parametrizationcheck_1Hz} shows a comparison of the accuracies 
of the measurement of the various PN parameters under different 
choices of the  parametrization schemes for a $(10,\, 100) M_{\odot}$ 
binary, located at a luminosity distance of 3 Gpc. The comparison uses
the RWF, with lower frequency cutoffs of 1 and 10 Hz, respectively. 

\begin{table*}
\centering
\caption{Accuracies of the measurement of various PN 
parameters (using the RWF and 1 Hz low-frequency cutoff) for 
(10, 100)$M_{\odot}$ binary located at a luminosity distance of 3 Gpc,
with  different choices of parametrization schemes. For each
entry, the number within parentheses is the factor by which the
accuracy will be reduced if a lower cutoff of 10 Hz is chosen instead
of 1 Hz.  For the fundamental pair we have chosen $(\psi_0, \psi_f)$ 
where $f$ can be any of $2,\,3,\,4,\,5l,\,6,\,6l,\,7$. In each case, the 
relative error in the  test parameter is listed in the third row.}
\label{Parametrizationcheck_1Hz}
\begin{tabular*}{1.00\textwidth}{@{\extracolsep{\fill}} c|c|c|c|c|c|c|c}

\hline
\hline
\multicolumn{8}{l}{{$(m_1,m_2) = (10,\,100)M_\odot $};\,\,\,{$f_{\rm s} = 1$ Hz};\,\,\,{$D_L=3$ Gpc};\,\,\,{waveform model: RWF}} \\ 
\hline
& $\psi_0$-$\psi_2$ & $\psi_0$-$\psi_3$ & $\psi_0$-$\psi_4$ & $\psi_0$-$\psi_{5l}$ & $\psi_0$-$\psi_6$ & $\psi_0$-$\psi_{6l}$ & $\psi_0$-$\psi_7$ \\
\hline
$\Delta \psi_0$/$\psi_0$     & ...  &    0.0015   (60) &   0.0015 (60)   &   0.0015 (60)  &    0.0015 (60)  &  0.0015 (60)  &   0.0015 (60)  \\

$\Delta \psi_f$/$\psi_f$     & ...  &    0.0092 (15)  &    0.010 (17)   &   0.017 (18)  &    0.043 (17) &   0.020 (19)  &   0.022 (19)  \\

$\Delta \psi_2$/$\psi_2$    & ...  &    0.027 (27)  &    0.027 (27)  &    0.027(27)  &    0.027(27)  &  0.027(27)  &   0.027(27)  \\

\hline

$\Delta \psi_0$/$\psi_0$     & 0.0010 (55) &  ...   &   0.0010 (55)  &    0.0010(55)   &   0.0010(55) &   0.0010(55) &    0.0010(55)  \\

$\Delta \psi_f$/$\psi_f$    & 0.0089 (13)  &   ...   &   0.020 (16) &    0.031(16)   &   0.082(16) &   0.037(16) &    0.042(16)  \\

$\Delta \psi_3$/$\psi_3$   & 0.0050 (42)   &   ...  &    0.0050 (42) &    0.0050 (42)  &   0.0050(42)  &  0.0050(42)  &   0.0050 (42)  \\

\hline

$\Delta \psi_0$/$\psi_0$   & 0.0011 (28) &    0.0011(28)  &    ... &     0.0011 (28)  &   0.0011(28) &   0.0011(28) &    0.0011(28)   \\

$\Delta \psi_f$/$\psi_f$     & 0.074(8)  &    0.15(8) &    ... &     0.25(8)  &    0.65(8) &   0.29(8) &    0.33(8)  \\

$\Delta \psi_4$/$\psi_4$   &  2.1 (8) &    2.1(8)   &   ... &     2.1(8)  &    2.1(8) &   2.1(8) &    2.1(8)  \\

\hline

$\Delta \psi_0$/$\psi_0$    &  0.00059 (77) &     0.00059 (77)&     0.00059(77) &     ...  &    0.00059(77) &   0.00059(77) &    0.00059(77)   \\

$\Delta \psi_f$/$\psi_f$    &  0.014(24)  &    0.026(23)  &    0.029(23) &     ...  &    0.12(23) &   0.052(23) &    0.058(23)   \\

$\Delta \psi_{5l}$/$\psi_{5l}$  & 0.056 (17)  &    0.056(17) &     0.056(17)  &   ...  &    0.056(17) &   0.056(17) &    0.056(17) \\

\hline

$\Delta \psi_0$/$\psi_0$    & 0.00054 (64) &    0.00054 (64)  &   0.00054 (64)  &   0.00054(64)  &    ... &    0.00054 (64)&    0.00054(64)   \\
 
$\Delta \psi_f$/$\psi_f$     & 0.0067 (21)&     0.013(20)  &    0.014(19)  &    0.021 (19) &    ... &   0.025(19) &    0.028(19)   \\
 
$\Delta \psi_6$/$\psi_6$  &  0.67(13)  &    0.67 (13) &    0.67(13)  &    0.67 (13)  &   ... &   0.67(13) &    0.67(13)   \\

\hline

$\Delta \psi_0$/$\psi_0$  &  0.00051(62)   &   0.00051 (62) &    0.00051(62)  &    0.00051(62) &     0.00051(62)   & ... &    0.00051 (62)  \\

$\Delta \psi_f$/$\psi_f$    &  0.0051(21) &     0.0096 (19)&     0.010 (19) &    0.016(19) &     0.042(19) &   ... &    0.021(18)    \\

$\Delta \psi_{6l}$/$\psi_{6l}$  & 0.17(13)   &   0.17 (13) &    0.17 (13)  &   0.17(13)  &    0.17(13) &  ... &    0.17(13) \\

\hline

$\Delta \psi_0$/$\psi_0$ & 0.00049 (59)  &    0.00049(59) &     0.00049(59)  &    0.00049(59)   &   0.00049(59) &   0.00049(59)  &   ...  \\

$\Delta \psi_f$/$\psi_f$ & 0.0046 (20) &    0.0087(18) &     0.0094(18)  &    0.014 (17)  &   0.038(18) &   0.017(17)  & ... \\

$\Delta \psi_7$/$\psi_7$ & 0.19(10)  &    0.19(10) &     0.19(10)  &    0.19(10)   &   0.19(10) &   0.19(10)  & ...  \\
\hline
\hline
\end{tabular*}
\end{table*}

From the table the following observations are evident: 
\begin{enumerate}
\item A comparison of the values in blocks symmetric across the 
principal diagonal one can compare the errors in the estimation 
of a particular parameter in the following two cases: once when 
the parameter is one of the {\it basic} variables and secondly when 
it is a {\it test} variable.  It is also clear that, in general, 
a parameter is determined more precisely when it is a basic parameter 
than when it is a test parameter. This is mainly 
because the basic variables bring new functional dependences via the 
rest of the phasing terms. 
\item The choice of the lower-order PN coefficient $\psi_2$ as a 
basic variable leads to a more precise test. 
\item When $\psi_0$ is one of the basic variables, the dispersion in the 
relative error of the {\it other basic} variable is least when the lowest-order PN coefficient $\psi_2$ is chosen as the {\it second basic} variable, 
\item An interesting case corresponds to the choice of $\psi_4$ as the 
basic variable which seems to allow for the best determination of $\psi_4$.
\end{enumerate}
As a result, although, in principle, one has the freedom of parametrizing 
the waveform in terms of {\it any} of the two PN coefficients, 
the choice of  ($\psi_0$ , $\psi_2$) as basic variables is the optimal one. 

The above question may be equivalently investigated by looking at
the volume of the three-dimensional ellipsoid corresponding to the three 
phasing coefficients which are involved in the test. We find that the 
smallest volume of the ellipsoid corresponds to the case where $\psi_0$ 
and $\psi_2$ are used as basic variables as compared to other combinations, 
for all test parameters except $\psi_4$ and $\psi_6$.  For these two 
parameters the volume is smaller when they are used as basic variables 
together with $\psi_0$. 
\subsubsection{The choice of angles}

In Sec.~\ref{TOG2-ET} we pointed out that the signal depends 
on four angular parameters ($\cos\theta$, $\phi$, $\psi$, $\cos\iota$) 
related to the source location and orientation but  that they were
chosen {\it arbitrarily} to be $\theta=\phi=\pi/6$, $\psi=\pi/4$, 
and $\iota=\pi/3$ in the present study. This is because for terrestrial detectors
and burst sources the angles could be considered constant. To quantify 
the effect of these angular dependences on the test we computed the 
relative error in  a particular PN parameter for 100 different 
realizations of these angular parameters.  

\begin{figure}[htbp!]
\includegraphics[width=0.42\textwidth,angle=0]{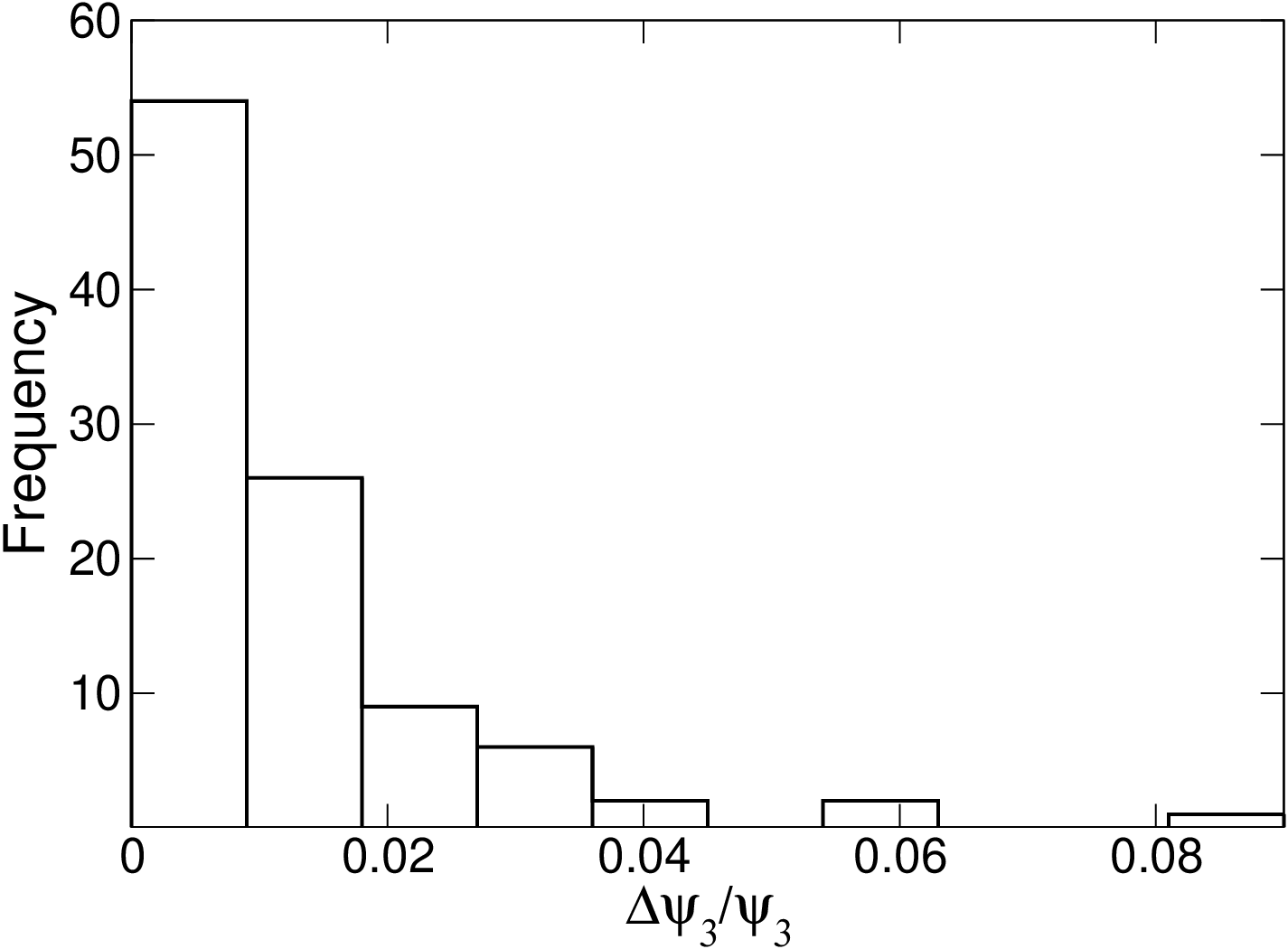}
\caption{Histogram for the relative error in the estimation of the 
parameter $\psi_3$ using 100 different realizations of angular 
parameters for a (10, 100)$M_{\odot}$ binary located at the luminosity 
distance of 3 Gpc. The low-frequency cutoff is 1 Hz and RWF has been used. }
\label{histogrampsi3}
\end{figure}

The result is plotted in Fig.~\ref {histogrampsi3}. From Figs.~\ref{histogrampsi3} and ~\ref{ErrVsmass_IMBH_0p1}, it is clear that the value of the 
relative error in the estimation of $\psi_3$ for a (10, 100)$M_{\odot}$
binary located at a luminosity distance of 3 Gpc is a typical value 
and, as was physically expected, the weak dependence on angles is 
a good approximation.
\section{Summary and concluding remarks}
\label{Summary}
In this paper we have studied the possibility of testing the 
theory of gravity 
within a well-defined subclass of ppE theories 
using GW observations of BBHs by a typical 
second generation GW interferometer (Advanced LIGO) and the 
plausible third generation GW interferometer (ET). Within
this subclass of theories for Advanced 
LIGO we have shown that GW observations of BBHs (in the 
range 11-110$M_\odot$ and at a luminosity distance 
of $ 300$ Mpc) can be used to estimate {\it only} the PN 
coefficient $\psi_3$ with fractional accuracy better than 6\% when 
the FWF is used (see Fig.~\ref{errvsmass_advligo}). Estimation 
of a PN coefficient with such an accuracy suggests that Advanced 
LIGO could indeed begin the era of strong-field tests of gravity. 
We have also compared the results for the FWF and RWF and shown 
that FWF reduces the errors by a factor of 3 to almost 100. 

We have also studied in detail the stellar mass and intermediate 
mass regimes of the compact binary source population in the ET 
sensitivity band, for 1 and 10 Hz lower cut off frequencies and 
compared the advantage of using the FWF model over the RWF model.
We find that the lower frequency cut off of 1 Hz plays a crucial 
role in testing GR with ET. 

For stellar mass binary coalescences (total mass $\leq 44 M_\odot$) 
as well as intermediate BH binaries, the lower cut off of 1 Hz improves 
the estimation of all PN parameters in the phasing formula.  For 
stellar mass binaries, the improvement in the estimation is between 
a factor of 2 to almost 20 when the RWF model is used. When the FWF model
is used, the improvements are typically between factors 2-10 (see 
Fig.~\ref{ErrVsmass_STMBH_0p1}).  

For intermediate mass binaries, which coalesce at lower frequencies, 
though the smaller lower cut off improves the parameter estimation, 
the errors associated with the measurement of various parameters 
are so large that the tests are not very interesting (see 
Fig.~\ref{ErrVsmass_IMBH_0p1}).  However, when total mass is less 
than about $100 M_\odot$, all the $\psi_k$'s are measured with 
relative errors less than unity, the most accurately determined 
parameters being $\psi_3$ and $\psi_{5l},$ which are determined with  
accuracies better than 10\%.  This seems to be the most interesting 
mass range for the proposed test in the ET band.  Though the use of 
the FWF does improve the estimation of various parameters, the test 
is less impressive since for astrophysically realistic event rates, 
we have to consider distances as large as 3 Gpc (as opposed to 300 
Mpc for the stellar mass case).  Thus, only if there is such an 
event very close by, can the test be performed very accurately.

It is worth bearing in mind that in addition to systematic effects
due to higher-order PN terms, various other systematic effects  
could offset the accuracy of the proposed test: 
\begin{enumerate}
\item If the components of the binaries have spins, the phasing 
coefficients are functions not only of the individual masses but 
also the spin parameters.  Further, if the binary is precessing 
(which would be the case when the spins are not aligned or 
antialigned with the orbital angular momentum vector 
of the binary), the waveforms will have a very different structure 
due to spin-induced modulations. To get a simple estimate of the 
effect of spins on the proposed test, we consider spinning but 
nonprecessing binaries. For such binaries, the effect of spin is to 
introduce  additional spin-dependent contributions in various phasing 
coefficients at and above 1.5PN.  The 1.5PN phasing coefficient in 
this case has an additional spin parameter $\beta$, which is a 
function of the individual spins of the binary and takes values 
$0\leq\beta\leq8.5$~\cite{CF94}. We found that for values of 
$\beta\geq6$, the bias in our estimate could be more than 100\%. 
This means that the presence of spins could significantly bias the proposed 
test for large values of the spin parameter. 
\item Another  effect is that of the orbital eccentricity, which 
we have ignored by assuming the binary's orbit to be quasicircular. 
As shown in Refs.~\cite{KKS95,YABW09}, orbital eccentricity will introduce 
additional phasing coefficients with completely different frequency 
dependences. It will need a careful study to assess how to 
incorporate the effect of eccentricity into our analysis, which 
we postpone to a future work. 
\item Lastly, since we have used 
PN inspiral waveforms, the neglect of  merger and ringdown effects
could also lead to further systematic errors.  By a proper choice 
of the domain of integration of the signal, we should be able to 
take care of it to some extent. A detailed study using some of the 
analytic parametrizations of numerical relativity waveforms (see 
Refs.~\cite{AjithNR07b,Damour:2009wj,PanNR09}) is planned as a follow up 
of this work.
\end{enumerate}
\vskip 0.4cm 
In this paper we have explored issues and indicated ways to test 
a class of theories of gravitation close to GR by using GW observations
in Advanced LIGO and the Einstein Telescope. The extension of these
results within a more general class of models like the ppE-framework
can be expected to  provide more general results in the future.  

To fully test our proposal one must mimic the whole exercise 
with mock data. One has to inject a {\it non-GR signal} into 
Gaussian background with a signal that differs from GR at 1.5 PN and 
higher orders by a certain degree. One would then need to extract, 
say, the first three parameters by an Markov chain Monte Carlo technique that employs
GR templates, and see if what we expect based on our
toy examples above holds good. We are currently exploring this exercise
but this exercise is quite compute intensive and goes beyond the scope of 
the present paper and so we defer its full discussion to a future 
publication.

\vskip 0.5cm

\begin{figure}[htbp!]\includegraphics[width=0.40\textwidth]{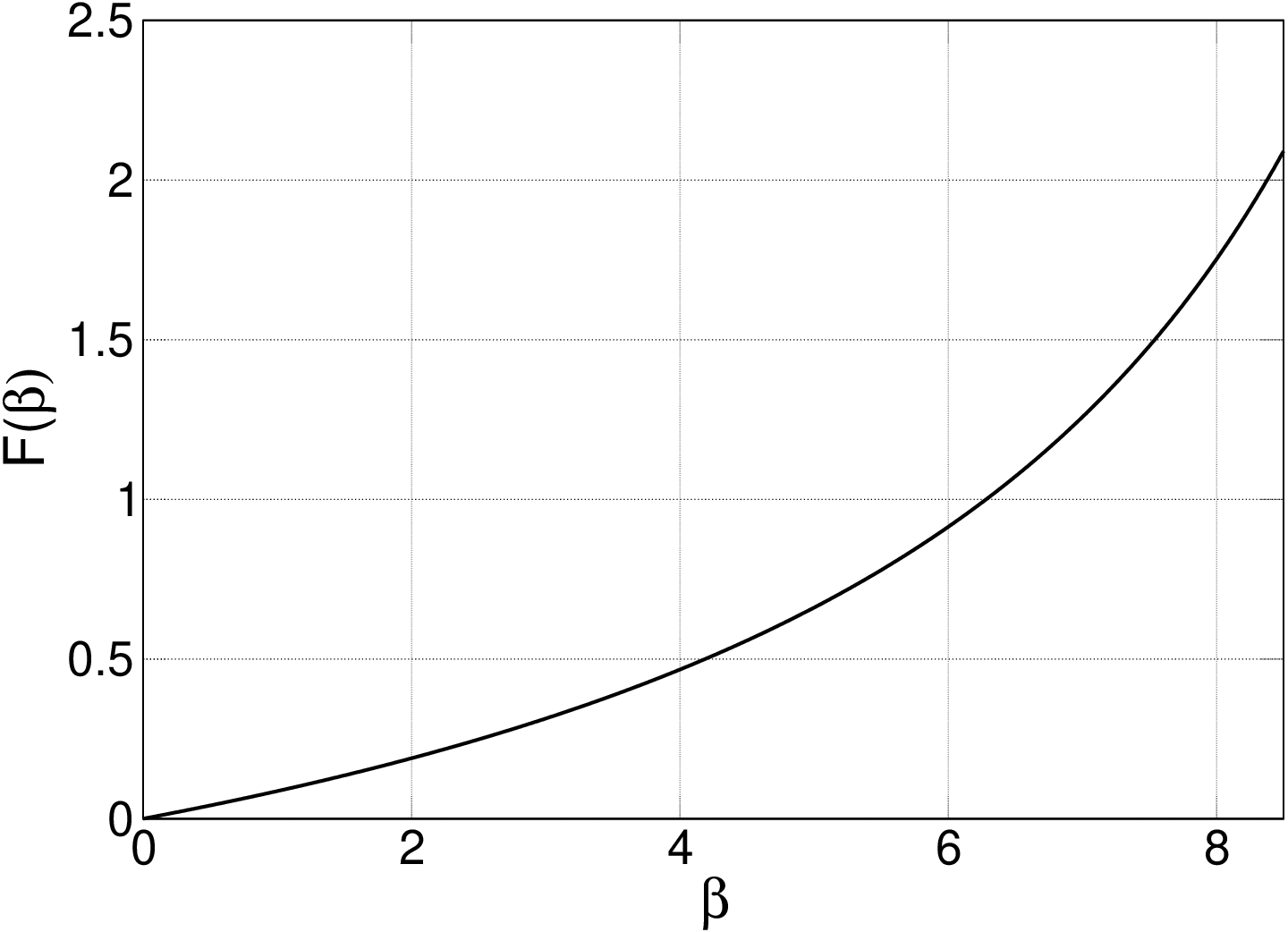}
\caption{This plot shows the variation of systematic bias due to spin $F(\beta)$ with the spin parameter $0\leq\beta\leq8.5$, where $F(\beta)$ is given by $F(\beta)=
4\beta(16\pi-4\beta)^{-1} $.}
\label{fig:bias}
\end{figure}

\begin{acknowledgments}
We thank Collin Capano (Syracuse) for providing the analytical 
fit for Advanced LIGO sensitivity curve. KGA thanks Clifford Will 
for discussions. K.G.A. acknowledges support by the National Science 
Foundation, Grant No.\ PHY 06--52448, the National Aeronautics 
and Space Administration, Grant No.\ NNG-06GI60G, and the Centre 
National de la Recherche Scientifique, Programme International 
de Coop\'eration Scientifique (CNRS-PICS), Grant No. 4396. BSS
was supported in part by PPARC Grant No. PP/B500731/1.
\end{acknowledgments}

\appendix
\section{Systematic effect due to spin}
We discuss the typical biases on our estimates due to the assumption
that the binary components are nonspinning. We demonstrate this, by taking
the 1.5PN phasing coefficient, where the spins first enter the phasing.
For convenience, we have assumed the spins of the binary are aligned with the
orbital angular momentum vector, in which case we can use the direct analytical
formula for the phasing coefficient.

As we mentioned earlier, the nonspinning 1.5PN phasing coefficient is given by
$\alpha_3^{\rm nonspin}=-16\pi$. The corresponding expression for spinning 
but nonprecessing binaries is $\alpha_3^{\rm spin}=-16\pi+4\beta$ where 
$\beta$ is a spin parameter which is a function of the spins of the individual 
components of the binary and whose value lies in the range $0\leq\beta\leq8.5$~\cite{CF94}. 
Thus the difference in the value of the 1.5PN coefficient due to spin is 
$\delta\alpha_3^{\rm spin}=\alpha_3^{\rm spin}-\alpha_3^{\rm nonspin}=4\beta$.
The bias in our estimates of ${\frac{\Delta \alpha_3}{\alpha_3^{\rm nonspin}}}$ 
is given by 
\begin{equation}
{\Delta \alpha_3 \over {\alpha_3}^{\rm spin}}-{\Delta \alpha_3 \over {\alpha_3}^{\rm nonspin}}={\Delta \alpha_3 \over {\alpha_3}^{\rm nonspin}}\times F(\beta)
\end{equation}
where $F(\beta)={\frac{4\beta}{(16\pi-4\beta)}}$ quantifies the bias in 
our estimate.

Figure~\ref{fig:bias} shows the plot of $F(\beta)$. As is obvious, the 
systematic bias due to spins could offset the estimation of $\alpha_3$ 
by more than 100\% for $\beta\geq6$.
\bibliography{ref-list}
\end{document}